\documentclass[a4paper,fleqn,usenatbib,twocolumn]{mnras}

\usepackage{graphicx}	
\usepackage{amsmath}	
\usepackage{amssymb}	
\usepackage{savesym}
\usepackage{natbib}
\usepackage{mathrsfs,dsfont,pifont}
\usepackage{multirow}
\usepackage{makecell}
\usepackage{cleveref}
\usepackage[skip=0pt]{caption}
\usepackage{float}
\usepackage{booktabs}
\usepackage{epstopdf}
\usepackage{romannum}
\usepackage{color}
\usepackage{hyperref}
\usepackage{verbatim}

\usepackage[none]{hyphenat}
\usepackage{soul}

\def\be{\begin{equation}}
\def\ee{\end{equation}}
\def\kms{{\rm \,km\,s^{-1}}}

\def\Ms{{\rm \,Ms}}

\def\Gpc{{\rm \,Gpc}}
\def\Mpc{{\rm \,Mpc}}
\def\kpc{{\rm \,kpc}}

\def\eV{{\rm \,eV}}
\def\keV{{\rm \,keV}}
\def\K{{\rm \,K}}
\def\msun{{\,M_\odot}}

\newcommand{\dd}{{\rm d}}

\usepackage{xcolor}

\title[Mapping ICM with High-resolution X-ray Spectroscopy]{Mapping the Intracluster Medium in the Era of High-resolution X-ray Spectroscopy}

\author[Congyao Zhang et al.]{Congyao Zhang,$^1$\thanks{E-mail: cyzhang@astro.uchicago.edu}
Irina Zhuravleva,$^1$
Maxim Markevitch,$^2$
John ZuHone,$^3$
Fran\c{c}ois Mernier,$^{2,4}$
\newauthor
Veronica Biffi,$^5$
\'Akos Bogd\'an,$^3$
Priyanka Chakraborty,$^3$
Eugene Churazov,$^{6,7}$
Klaus Dolag,$^{8,6}$
\newauthor
Stefano Ettori,$^{9,10}$
William R. Forman,$^3$
Christine Jones,$^3$
Ildar Khabibullin,$^{8,6,7}$
Caroline Kilbourne,$^2$
\newauthor
Ralph Kraft,$^3$
Erwin T. Lau,$^3$
Sheng-Chieh Lin,$^{11}$
Daisuke Nagai,$^{12}$
Dylan Nelson,$^{13}$
\newauthor
Anna Ogorza{\l}ek,$^{2,4}$
Elena Rasia,$^{5,14}$
Arnab Sarkar,$^{15}$
Aurora Simionescu,$^{16,17,18}$
Yuanyuan Su,$^{11}$
\newauthor
Mark Vogelsberger,$^{15}$
and
Stephen Walker$^{19}$
\\ \\
$^1$~Department of Astronomy and Astrophysics, The University of Chicago, Chicago, IL 60637, USA \\
$^2$~NASA Goddard Space Flight Center, Greenbelt, MD 20771, USA \\
$^3$~Center for Astrophysics \ding{120} Harvard \& Smithsonian, 60 Garden Street, Cambridge, MA 02138, USA \\
$^4$~Department of Astronomy, University of Maryland, College Park, MD 20742, USA \\
$^5$~INAF - Osservatorio Astronomico di Trieste, via Tiepolo 11, I34131 Trieste, Italy \\
$^6$~Max Planck Institute for Astrophysics, Karl-Schwarzschild-Str. 1, D-85741 Garching, Germany  \\
$^7$~Space Research Institute (IKI), Profsoyuznaya 84/32, Moscow 117997, Russia \\
$^8$~University Observatory Munich, Scheinerstr 1, D-81679 Munich, Germany \\
$^9$~INAF, Osservatorio di Astrofisica e Scienza dello Spazio, via Piero Gobetti 93/3, 40129 Bologna, Italy \\
$^{10}$~INFN, Sezione di Bologna, viale Berti Pichat 6/2, 40127 Bologna, Italy \\
$^{11}$~University of Kentucky, 505 Rose street, Lexington, KY 40506, USA \\
$^{12}$~Physics Department, Yale University, New Haven, CT 06511, USA \\
$^{13}$~Universitat Heidelberg, Zentrum f\"{u}r Astronomie, Institut f\"{u}r theoretische Astrophysik, Albert-Ueberle-Str.~2, \\  \quad 69120 Heidelberg, Germany \\
$^{14}$~Institute for Fundamental Physics of the Universe, via Beirut 2, 34151, Trieste, Italy \\
$^{15}$~Department of Physics, Kavli Institute for Astrophysics and Space Research, Massachusetts Institute of Technology, \\  \quad Cambridge, MA 02139, USA\\
$^{16}$~SRON Netherlands Institute for Space Research, Niels Bohrweg 4, NL-2333 CA Leiden, the Netherlands\\
$^{17}$~Leiden Observatory, Leiden University, PO Box 9513, NL-2300 RA Leiden, the Netherlands\\
$^{18}$~Kavli Institute for the Physics and Mathematics of the Universe, University of Tokyo, Kashiwa 277-8583, Japan\\
$^{19}$~University of Alabama in Huntsville, Department of Physics and Astronomy, Huntsville, AL 35899
\vspace{-30pt}
}

\date{}

\pubyear{2023}

\begin{document}
\label{firstpage}
\pagerange{\pageref{firstpage}--\pageref{lastpage}}
\maketitle

\begin{abstract}
\vspace{-8pt}

High-resolution spectroscopy in soft X-rays will open a new window to map multiphase gas in galaxy clusters and probe physics of the intracluster medium (ICM), including chemical enrichment histories, circulation of matter and energy during large-scale structure evolution, stellar and black hole feedback, halo virialization, and gas mixing processes. An eV-level spectral resolution, large field-of-view, and effective area are essential to separate cluster emissions from the Galactic foreground and efficiently map the cluster outskirts. Several mission concepts that meet these criteria have been proposed recently, e.g., LEM, HUBS, and Super\,DIOS. This theoretical study explores what information on ICM physics could be recovered with such missions and the associated challenges. We emphasize the need for a comprehensive comparison between simulations and observations to interpret the high-resolution spectroscopic observations correctly. Using Line Emission Mapper (LEM) characteristics as an example, we demonstrate that it enables the use of soft X-ray emission lines (e.g., O\,VII/VIII and Fe-L complex) from the cluster outskirts to measure the thermodynamic, chemical, and kinematic properties of the gas up to $r_{200}$ and beyond. By generating mock observations with full backgrounds, analysing their images/spectra with observational approaches, and comparing the recovered characteristics with true ones from simulations, we develop six key science drivers for future missions, including the exploration of multiphase gas in galaxy clusters (e.g., temperature fluctuations, phase-space distributions), metallicity, ICM gas bulk motions and turbulence power spectra, ICM-cosmic filament interactions, and advances for cluster cosmology.

\end{abstract}

\begin{keywords}
galaxies: clusters: intracluster medium -- methods: numerical -- techniques: imaging spectroscopy -- X-rays: galaxies: clusters
\vspace{-20pt}
\end{keywords}


\section{Introduction} \label{sec:introduction}

The intracluster medium (ICM) is a gaseous component of galaxy clusters that fills the space between galaxies, contributing the largest fraction of the cluster's baryonic mass. It is predominantly composed of low-density but remarkably hot ($\sim 10^7 - 10^8$ K) weakly-magnetized plasma, emitting X-rays via bremsstrahlung and lines of heavy elements (see, e.g., \citealt{Kravtsov2012,Vikhlinin2014,Walker2019} for reviews).

A number of processes that drive cluster growth (e.g., smooth accretion, mergers) and evolution of the galaxies (e.g., feedback, gas stripping, and mixing) are imprinted on the ICM, allowing detailed studies with X-ray and sub-mm/Sunyaev-Zel’dovich (SZ) observations. In recent years, the gas thermodynamic and chemical properties of many clusters have been measured in the past years from core regions (e.g., by \textit{Chandra}, \textit{XMM-Newton}) up to virial radii (e.g., \textit{Suzaku}, SRG/eROSITA, and \textit{Planck}). However, low spectral resolution of current CCD-type X-ray detectors prevents direct gas velocity measurements through the broadening and shift of emission lines.\footnote{With the Reflection Grating Spectrometers (\textit{RGS}) on \textit{XMM-Newton}, it is possible to probe gas dynamics within the compact innermost regions of some galaxy clusters, group and massive galaxies \citep[e.g.,][]{Sanders2010,dePlaa2012,Pinto2015,Ogorzalek2017}.} Only in some special cases, CCD-type detectors could be pushed to their limit to provide tentative detections of the ICM bulk motions \citep[e.g.,][]{Tamura2014,Liu2015,Sanders2020,Gatuzz2022}. The lack of sufficient energy resolution also prevents robust measurements of different gas phases in the ICM and their coupling, and abundances of various elements beyond Fe in most cases (see \citealt{Mernier2018} for a review).

Only recently, the \textit{Hitomi} satellite, with an X-ray microcalorimeter on board, mapped the velocity field and measured metal abundances in the central region of the Perseus cluster with an unprecedented spectral resolution of $\simeq 5\eV$ \citep{Hitomi2016,Hitomi2017}. As its successor, the recently-launched \textit{XRISM} observatory will soon observe many extended X-ray sources with similar spectral capabilities \citep{XRISM2020,Ezoe2021}, resolving individual lines in spectra of galaxy clusters, providing kinematic properties of the ICM and allowing multi-component gas decomposition (both in temperature and velocity). Given a modest effective area of the telescope, {\it XRISM} will mainly focus on the brightest cluster regions within $r_{2500}$\footnote{The $r_{\Delta}$ represents the radius within which an average mass density is $\Delta$ (e.g., $2500,\ 500,\ 200$) times the critical density of the Universe at the cluster redshift.} with $\simeq1'$ angular resolution.

Resolving X-ray spectra beyond the central cluster regions ($\gtrsim r_{2500}$) is crucial for understanding chemical enrichment histories, global energy transfer during large-scale structure evolution, how material is brought from the intergalactic medium (IGM) into the ICM and gets mixed, gaseous boundaries and halo virialization, and cluster cosmology. From a theoretical point of view, understanding ICM physics in the cluster outskirts, especially the IGM-ICM transition regions, is a complex multi-scale problem, spanning $3-4$ orders of magnitude, from Mpc to sub-kpc (if not smaller, considering the effects of magnetic fields). These outer regions are poorly resolved in current cosmological simulations, given the extremely low gas number density ($\lesssim10^{-4}\,{\rm cm}^{-3}$). Therefore, properly capturing non-linear processes, e.g., small-scale turbulence, multiphase gas interactions, and discontinuities, is non-trivial.

\begin{figure}
\centering
\includegraphics[width=0.9\linewidth]{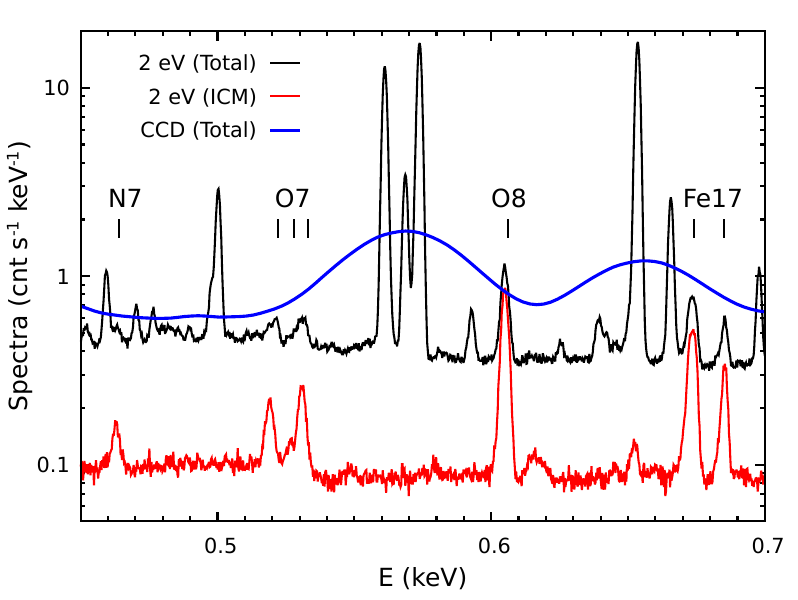}
\caption{
High-resolution X-ray spectroscopy is crucial for separating diffuse components of the Milky Way and the ICM emission from the galaxy cluster outskirts. The black and red curves show spectra of a cluster in the annulus 1 - 1.1$r_{500}$ (ICM temperature $\simeq1.5\keV$) at $z=0.08$, including all background components and the ICM-only component, respectively, simulated with the $2\eV$ resolution. Multiple strong emission lines from the cluster (vertical dashes) stand out above the background. For comparison, the blue curve shows the total spectrum with optimistic CCD resolution ($50\eV$), for which separating the ICM from the background is impossible (see Section~\ref{sec:introduction}).
}
\label{fig:ccd}
\end{figure}

Detailed observations of the ICM in cluster outskirts through mapping thermodynamic, chemical, and kinematic properties will shed light on the missing elements of the current theoretical picture and stimulate the development of more physically motivated models. For this goal, one needs a telescope with a large grasp - the product of the effective area and field-of-view (FOV) of a telescope and high spectral resolution. A large effective area is required to collect sufficient photons with reasonable exposure times and a large FOV to map extended outer regions efficiently.

\begin{figure}
\centering
\includegraphics[width=0.9\linewidth]{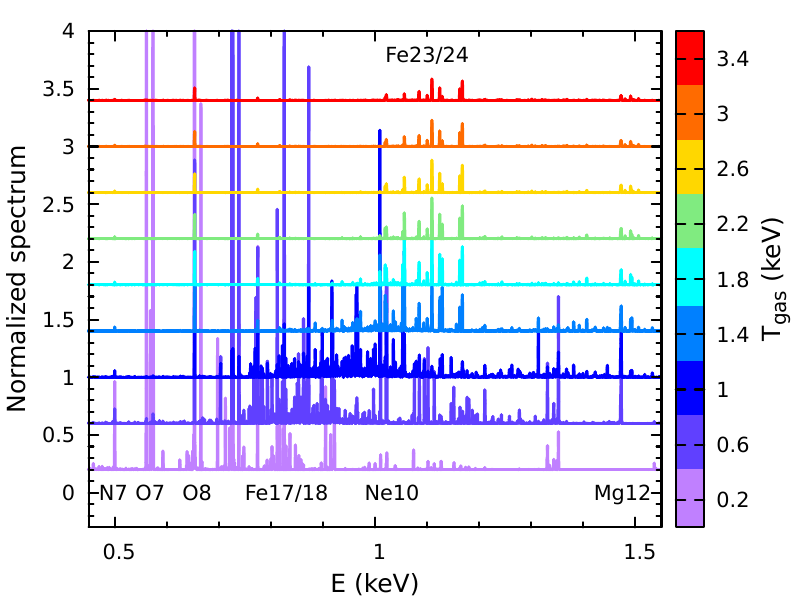}
\caption{Spectral template as a function of gas temperature $T_{\rm gas}$ (colours) from the APEC model, assuming solar metallicity. In the soft X-ray band, emission lines are extremely sensitive to various temperatures below $\simeq2\keV$, providing a direct opportunity to investigate multiphase gas in cluster outskirts (see Section~\ref{sec:introduction}).}
\label{fig:apec_spec}
\end{figure}

Intermediate and low-mass clusters ($\lesssim5\times10^{14}\msun$) dominate the cluster mass function. Their typical gas temperatures are $\simeq1-2\keV$ beyond $r_{500}$, corresponding to the peak of their X-ray emission in a soft X-ray band, at energies below $\simeq2\keV$.
However, at these low energies, the X-ray emission is dominated by a signal from the Milky Way that shows a forest of emission lines, with the strongest ones O\,VII and O\,VIII at the rest energies $\simeq0.57\keV$ and $0.65\keV$, respectively. In order to detect emission lines from the ICM distinguished clearly from the Galactic foreground, the targets have to be selected within specific redshift ranges, ideally $z\simeq0.06-0.1$\footnote{Corresponding to $\simeq0.24\Gpc^3$ comoving volume, where it is expected to find a few thousand clusters with $M_{200}>10^{14}\msun$ \citep[e.g.,][]{Bocquet2016}.}, such as Abell~3112 and 2597 \citep[e.g.,][]{Nevalainen2003,Tremblay2018}, and the X-ray spectral resolution of the instrument should be at least $\simeq 1-2\eV$ for the sake of accurate line profile measurements. This is illustrated in Fig.~\ref{fig:ccd}. The redshifted strong emission lines from the ICM (e.g., Fe\,XVII and O\,VIII) lie prominently above the background. They provide valuable information on gas temperature, metal abundances, velocity dispersion, and bulk velocity through line ratios, broadening, and shift measurements. Emission lines in soft X-rays are extremely sensitive to gas temperatures below $\simeq2\keV$, see an illustration in Fig.~\ref{fig:apec_spec} based on the Astrophysical Plasma Emission Code (APEC) model \citep{Smith2001}. Such features provide a unique opportunity to separate multiple gas phases, particularly important in the cluster outskirts, where projection effects mix the hot, volume-filling ICM, infalling substructures and their stripped medium, and/or penetrating filaments.

Recently, several mission concepts that meet all of the above criteria have been proposed, including Line Emission Mapper (LEM; see \citealt{Kraft2022}), Hot Universe Baryon Surveyor (HUBS; see \citealt{Cui2020,Bregman2023}), and Super Diffuse Intergalactic Oxygen Surveyor (Super\,DIOS; see \citealt{Yamada2018,Ohashi2018}) - all are soft X-ray missions with $1-2\eV$ spectral resolution.\footnote{Athena/NewAthena is another X-ray mission concept that covers a broader energy band ($0.2-12\keV$) and enables spatially resolved high-resolution X-ray spectroscopy with its X-IFU \citep{Barret2020}. It shows strong capabilities to map ICM velocities in massive clusters through the bright $6.7\keV$ Fe\,XXV~K$\alpha$ line (e.g., \citealt{Biffi2013,Roncarelli2018}; see also \citealt{Cucchetti2018,Mernier2020} for the measurement of metal abundance within $r_{500}$). However, limited by the small grasp of the X-IFU, observations would require a prohibitive time investment to cover the entire cluster outskirts.}
It is important to explore quantitatively what opportunities such missions provide in the studies of ICM, in particular, what can be learnt about gas kinematics and thermodynamics, chemical composition, and gas multiphaseness, especially in complex and dynamic cluster outer regions. This is the primary scope of this paper, i.e., to provide a theoretical guide on developing ICM/IGM science drivers for future missions. Without loss of generality, we will focus on LEM characteristics -- an X-ray Probe Mission Concept for the 2030s, equipped with a large microcalorimeter array covering a $30'\times30'$ FOV with $15''$ pixel size,\footnote{We used a $32'\times32'$ LEM FOV in all our calculations, which was recently updated during the manuscript preparation. This update does not affect any of our conclusions.} $1600\,{\rm cm}^2$ effective area at $0.5\keV$, and $2\eV$ energy resolution and a central sub-array ($5'\times5'$) with $0.9\eV$ resolution \citep{Kraft2022}, when producing mock data based on cosmological simulations. In addition to exploring what information could be recovered and the challenges, a comprehensive comparison between simulations and observations will be carried out to highlight the complexities of interpreting real data and understanding their underlying physics. As shown below, this is a crucial step for correctly bridging simulations and observations. In general, our methodology and most of the conclusions are applicable to any of the telescopes mentioned above.

This paper is organized as follows. Section~\ref{sec:sample} describes the cluster samples used in this study and their gas properties directly related to the X-ray spectra. In Section~\ref{sec:mock}, we present the main method of generating and analysing mock observational spectra. In Section~\ref{sec:results}, we elucidate six key scientific objectives in detail for the mission and their implications for understanding cluster physics. In Section~\ref{sec:conclusions}, we summarize our conclusions.

\section{TNG300 Cluster Sample and ICM Properties} \label{sec:sample}

\begin{table*}
\centering
\begin{minipage}{\linewidth}
\centering
\caption{Properties of clusters selected from the TNG300-1 sample for detailed studies in this work (see Section~\ref{sec:sample}). }
\label{tab:cluster_sample}
\begin{tabular}{lcccccc}
  \hline
  Name\footnote{The cluster's name used in the paper.} &
  TNG-ID\footnote{The ID of the cluster given in TNG300-1's group catalog. }    &
  $M_{200}\ (10^{\rm 14}\msun)$  &
  $r_{500}\ (\rm Mpc)$           &
  $r_{200}\ (\rm Mpc)$ &
  Status &
  $z$\footnote{The assumed redshift of the cluster for LEM mock data, though our clusters are all from the TNG300-1 simulation's $z=0$ snapshot. It is selected so that the major emission lines of the ICM can be well separated from the Galactic foreground and the virial region of the cluster ($\le r_{200}$) can be covered by one LEM FOV. }
\\ \hline
 CL-RS & 446262  &  $0.95$  & $0.64$  & $0.97$  & fully-relaxed  & 0.06 \\
 CL-RM & 180645  &  $2.8$   & $0.92$  & $1.38$  & fully-relaxed  & 0.08 \\
 CL-RL & 55060   &  $6.3$   & $1.19$  & $1.81$  & relaxed   & 0.12 \\
 CL-U & 150265  &  $3.3$   & $0.95$  & $1.46$  & unrelaxed & 0.08 \\
\hline
\vspace{-18pt}
\end{tabular}
\end{minipage}
\end{table*}

In this section, we briefly describe the simulated clusters used in this study and characterize their physical properties directly relevant to the high-resolution X-ray spectra.

Our cluster sample comes from the TNG300-1 cosmological simulation, a simulation set with the largest box size ($\simeq300\Mpc$) in the IllustrisTNG project \citep{Pillepich2018,Springel2018,Naiman2018,Marinacci2018,Nelson2018,Nelson2019}, including $\simeq340$ galaxy clusters (total bound mass $\geq10^{14}\msun$) at redshift zero. Among these clusters, we select $30$ most relaxed ones based on the following criteria: the clusters show visually smooth and spherical/near spherical gas distributions in their cores ($\lesssim r_{500}$) and no prominent substructures within $r_{200}$ based on their projected gas distributions.
The statistical properties of the full and relaxed cluster samples are compared in this section.

For the spectral analysis, we mostly focus on four clusters from the TNG300 sample to demonstrate the capability and robustness of high-resolution X-ray spectroscopy to understand the ICM physics (see Table~\ref{tab:cluster_sample}): three relaxed ones, Virgo- to Perseus-like clusters in terms of their total masses, and one unrelaxed system (CL-U) with deeply penetrated filaments. The cluster CL-RM with intermediate mass is one of the most relaxed clusters in the TNG simulations, representing an ideal target for investigating cluster hydrostatic mass bias. CL-RS, also fully relaxed, is similar to a fossil group/cluster \citep[e.g.,][]{Jones2003,Milosavljevic2006}, isolated from prominent large-scale structures (e.g., halos, filaments; see Fig.~\ref{fig:hist_vr2}) on the $\sim10\Mpc$ scale. CL-RL is a relatively perturbed cluster in our relaxed cluster sample, revealing velocity substructures within $r_{500}$.

Current X-ray observations suggest a flat metallicity radial profile of the ICM at $\gtrsim0.5r_{500}$ \citep[e.g.,][]{Urban2017,Ghizzardi2021}. There are, however, still large uncertainties in the cosmological simulation predictions, sensitive to the sub-grid models and their parameters. For the feasibility study, we assume that our clusters have uniform metallicity in units of \textsc{angr} solar abundances ($Z=0.2$; see \citealt{Anders1989}), throughout their volumes, unless otherwise stated. It provides a simple, physically-motivated baseline model to characterize plausible errors and biases in our mock observational measurements (see more discussions in Section~\ref{sec:results:metal}).

\subsection{Gas radial velocity histograms} \label{sec:sample:vr}

\begin{figure*}
\centering
\includegraphics[width=0.9\linewidth]{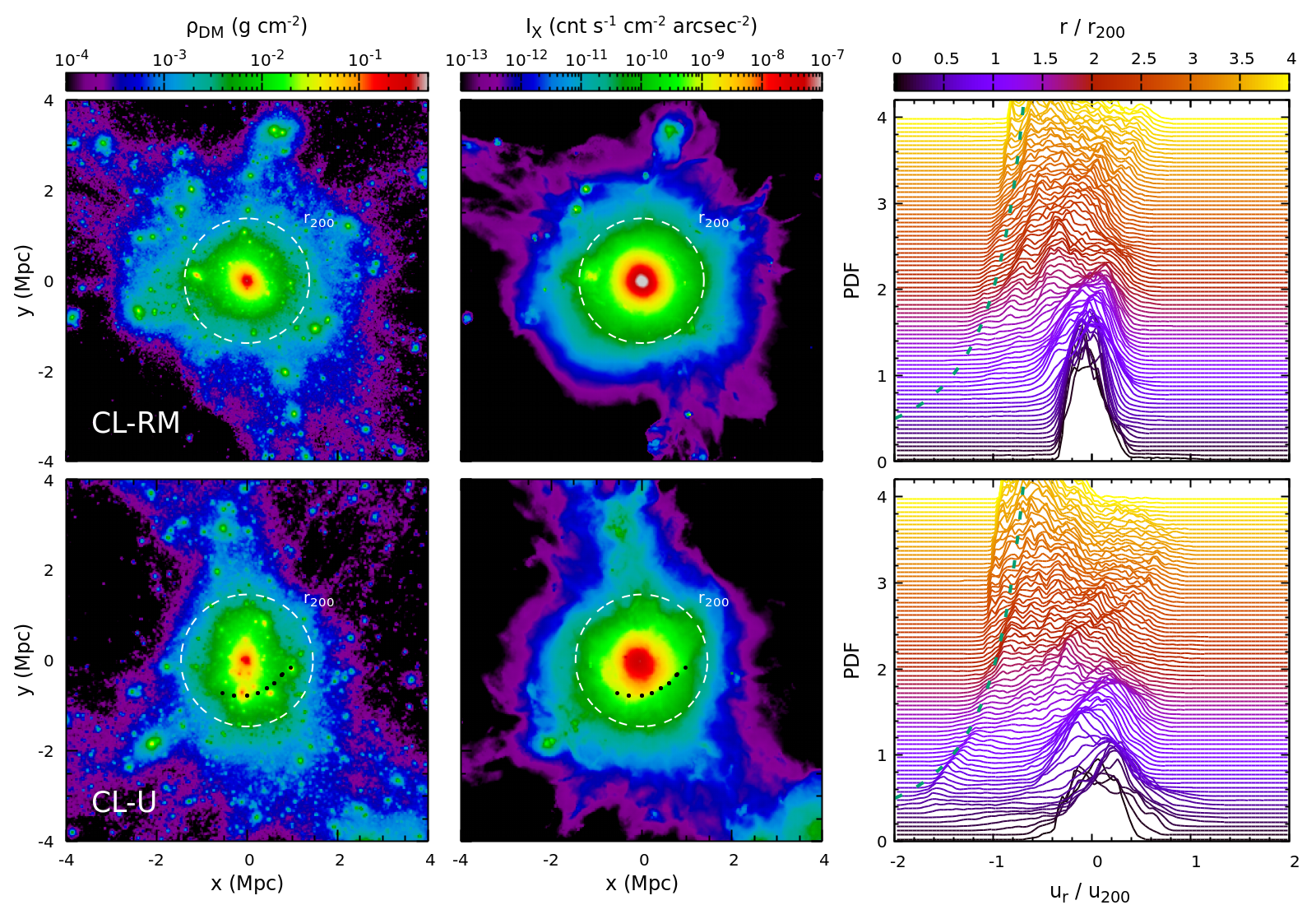}
\caption{Examples of a relaxed (the top panels) and a perturbed with deeply-penetrated filaments (the bottom panels) clusters (see Fig.~\ref{fig:hist_vr2} for the other two relaxed clusters listed in Table~\ref{tab:cluster_sample}). The left and middle panels show dark matter surface density and broad-band X-ray surface brightness ($0.5-2\keV$), including all dark matter particles/gas cells within $4r_{200}$. The white dashed circles indicate cluster $r_{200}$. 
The black dotted line marks a Mpc-scale bow shock identified in CL-U. The right panels show histograms of gas radial velocities in narrow radial shells, providing a convenient way to identify infalling substructures (e.g., subhalos, filaments). The green dashed line illustrates the free-fall velocity profile of the system, i.e., $u_{\rm r}/u_{200}=-\sqrt{2r_{200}/r}$. Penetrating filaments appear as ripples and bumps on the negative side (towards the cluster centre) of the velocity distributions (see Section~\ref{sec:sample:vr}). }
\label{fig:hist_vr}
\end{figure*}

In this study, we focus on two types of cluster: (1) relaxed ones and (2) perturbed ones with deeply penetrated filaments largely driven by mergers. They represent two key states of clusters in the hierarchical structure formation scenario. Analysing histograms of gas radial velocities in radial shells provides a robust and convenient way to identify them from a large cluster sample. This is illustrated in Fig.~\ref{fig:hist_vr}. Each curve in the right panels shows a volume-weighted ($m_{\rm gas}/\rho_{\rm gas}$) histogram of gas radial velocity at a given radius scaled by $u_{200}=\sqrt{GM_{200}/r_{200}}$, where $G$ is the gravitational constant, $ m_{\rm gas}$ and $\rho_{\rm gas}$ are mass and density of gas cells. The ripples and bumps emerging in the distribution highlight visually infalling substructures (e.g., small halos, filaments) and merger-driven outflows.

In the top panels, we show the relaxed cluster CL-RM. Its radial velocity distribution is smooth and narrow within $\simeq1.5r_{200}\,(\simeq r_{100}$), demonstrating the well-relaxed state of the cluster. At larger radii, substructures start to appear (e.g., ripples and asymmetries). Most of them have negative radial velocities, indicating that they are falling towards the cluster centre. While we see subhalos in the dark matter surface density and broad-band\footnote{Note that, we use the energy range $0.5-2\keV$ in our X-ray predictions throughout this paper (referred to as `broad-band' hereafter), including both surface brightness and X-ray-weighted quantities. We have tested a wider range $0.2-2\keV$, which does not affect any of our conclusions.} X-ray surface brightness within $r_{200}$, they are mostly projected from larger radii.

The bottom panels of Fig.~\ref{fig:hist_vr} show the unrelaxed CL-U. It is one of the TNG300 clusters we identified that has large-scale filaments deeply penetrated into the ICM, similar to those discussed by \citet{Zinger2016}. In both dark matter and X-ray maps, one can see the primary filament along the $y$-axis. The cluster is also elongated along the same direction in projection. The filament gas inflows all the way down to $0.5r_{200}$ and approximately follows the free-fall velocity profile $u\propto\sqrt{1/r}$ assuming a simple point-mass gravitational potential (green dashed line in the right panel). In 3D, the filament has a $\simeq30-45^\circ$ inclination angle from the line-of-sight (LOS; i.e., $z$-direction in the simulation). In Section~\ref{sec:results:filaments}, we will see that such a geometric configuration is ideal for detecting large-scale cosmic flows inside the ICM driven by filaments in observations.

It is worth noting that CL-U is a merging system. An infalling subcluster has just passed the core and is located near the inner end of the filament, leading the filament penetration into the ICM. The bow shock driven by the subcluster is marked by the black dotted lines. The associated gas radial inflow shown in the bottom right panel is, therefore, a mix of the merger-driven wake and filament. Since most of the mergers occur along the filaments, it is non-trivial to separate the diffuse gas of the filament and embedded subhalos within the filament. Therefore, in this study, we refer to these two components collectively as a `filament'. In the TNG300 cluster sample, we identified $\sim10$ clusters (out of $\simeq340$) with clear large-scale radial inflows reaching at least $r_{200}$, similar to CL-U. They all show signs of ongoing mergers (e.g., multiple subhalos near the cluster core, bow shocks) taking place close to the filamentary directions. Mergers likely play an important role in enabling filaments to penetrate deeply into the ICM. However, more quantitative studies on this topic are beyond the scope of this paper. We postpone them to our future work.


\subsection{Gas velocity dispersion}  \label{sec:sample:vsigma}

\begin{figure}
\centering
\includegraphics[width=0.9\linewidth]{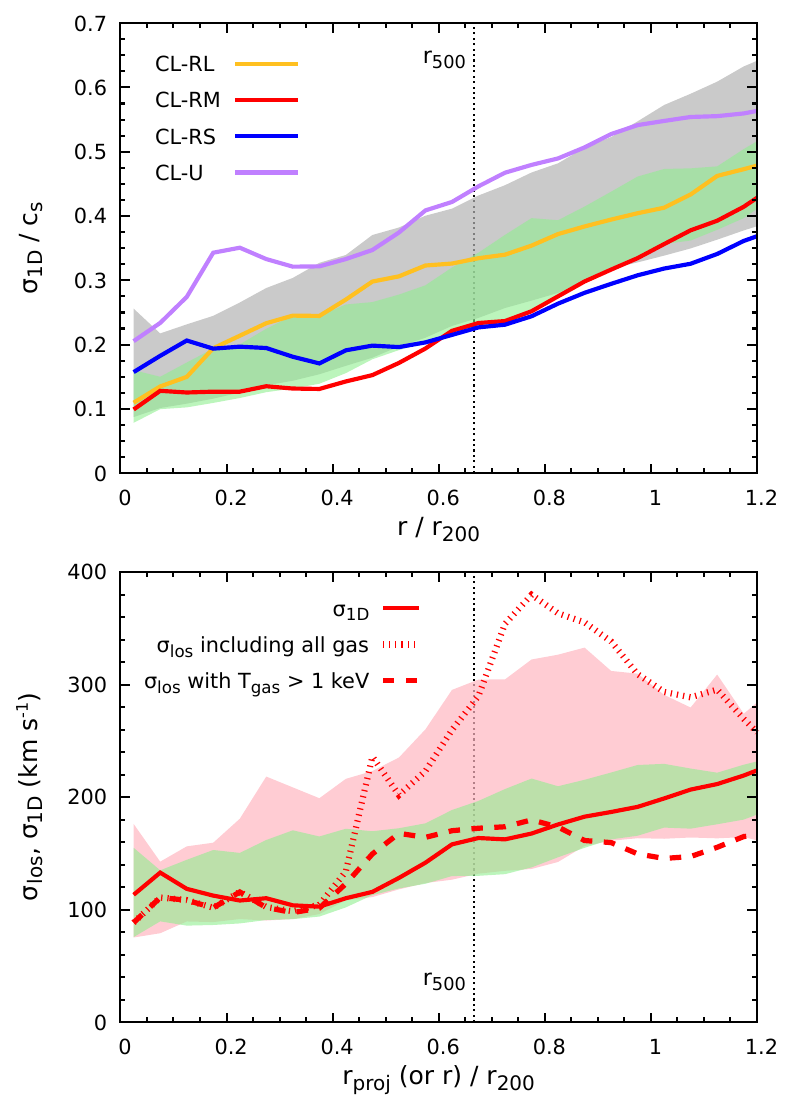}
\caption{\textit{Top panel:} radial profiles of gas 1D-velocity dispersion $\sigma_{\rm 1D}(r)$ in units of the local sound speed. The bands indicate the 10th and 90th percentiles of the dispersion over the whole cluster sample (grey) and the relaxed cluster subsample (green). The solid lines show the profiles of four individual clusters listed in Table~\ref{tab:cluster_sample}. The vertical dashed line marks $r_{500}$. \textit{Bottom panel:} a comparison between $\sigma_{\rm 1D}$ (green band and red solid line, same as in the top panel but not normalized to the sound speed) and broad-band X-ray-weighted gas LOS velocity dispersion $\sigma_{\rm los}$ (pink band, and red dashed and dotted lines) of the relaxed clusters. In particular, the lines compare the results for CL-RM. The dotted and dashed lines take into account all gas cells and those that are only hotter than $1\keV$, respectively. The cold clumps and small substructures make $\sigma_{\rm los}$ strongly deviate from $\sigma_{\rm 1D}$ even within $r_{500}$ due to the projection effect (see Section~\ref{sec:sample:vsigma}). }
\label{fig:tng_sigma}
\end{figure}

Fig.~\ref{fig:tng_sigma} shows radial profiles of characteristic gas 1-dimensional (1D) velocity dispersion $\sigma_{\rm 1D}(r)$ normalized by the local sound speed $c_{\rm s}(r)$ of our cluster sample,
\be
\sigma_{\rm 1D}(r) = \sqrt{\frac{1}{3}\sum_{i=x,y,z}{\sigma_{i}(r)^2}},
\label{eq:sigma1d}
\ee
where $\sigma_{i}$ is the mass-weighted velocity dispersion along $x$, $y$, and $z$-directions at the given radius. The shaded regions indicate the 10th and 90th percentile range of $\sigma_{\rm 1D}$ over the sample. Note that large-scale bulk motions are not filtered out in the estimation \citep[c.f.,][]{Vazza2018} so that we can directly compare these results with observations. Mergers and their associated infalling substructures may significantly boost the velocity dispersion. To illustrate such an effect, we compare the velocity dispersion profiles over the full TNG300-cluster sample (grey) and the subgroup of relaxed clusters (green). Relaxed clusters show smaller $\sigma_{\rm 1D}$ scatter by a factor of $\sim2$ at all radii. The average Mach number of their gas motions ranges from $\simeq0.1$ in the central region to $\simeq0.4$ near $r_{200}$. The latter corresponds to a non-thermal pressure fraction $\sim20\%$, generally consistent with those reported in the literature \citep[e.g.,][]{Nelson2014,Biffi2016,Zhuravleva2023}. Interestingly, two samples show comparable lower boundaries (10th percentile) of the velocity dispersion, reflecting the level of gas random motions in the most quiescent regions of a cluster. Velocity dispersion profiles of the four selected clusters are overlaid in the top panel (solid lines). CL-RM and CL-RS are clearly among the most relaxed clusters in our sample.

The bottom panel of Fig.~\ref{fig:tng_sigma} compares $\sigma_{\rm 1D}$ of our relaxed clusters with a broad-band X-ray-weighted LOS velocity dispersion $\sigma_{\rm los}$ that includes all gas cells within $4r_{200}$. Cold clumps\footnote{In the paper, we use the term `clumps' to refer generally to cold, dense, small gaseous structures inside a cluster or along its LOS, which may or may not associate with any galaxies or subhalos.} make $\sigma_{\rm los}$ significantly deviate from $\sigma_{\rm 1D}$ even within $r_{500}$ due to the projection effect. Taking CL-RM as an example, a large difference between $\sigma_{\rm los}$ and $\sigma_{\rm 1D}$ starts from $\sim 0.5r_{200}$ (see the red dotted and solid lines), although the cluster is well relaxed and lacks prominent substructures within $r_{200}$ (see Fig.~\ref{fig:hist_vr}). To exclude cold clumps, we take into account only the hot gas with $T_{\rm gas}>1\keV$, which excludes $\lesssim30$ per cent gas mass near $r_{200}$ but $\simeq70$ per cent of X-ray flux along the LOS -- a simple and robust way to remove the bulk clumps' contributions (see Fig.~\ref{fig:thist_cumu_180645} and Appendix~\ref{sec:appendix}). It corresponds to the red dashed line, appearing close to $\sigma_{\rm 1D}$. The deviation near $r_{200}$ is mainly caused by a large substructure in the east of the cluster (see Fig.~\ref{fig:hist_vr}). This suggests that the observable $\sigma_{\rm los}$ is a robust tracer of the true 1D velocity dispersion (in 3D) of the bulk gas component in relaxed clusters if it is possible to separate spatially and/or spectroscopically the cold gas contribution from the hot ICM. In Section~\ref{sec:results:vsigma}, we demonstrate the capability of high-resolution X-ray spectroscopy to recover $\sigma_{\rm 1D}$ in relaxed clusters.

\subsection{Temperature distribution} \label{sec:sample:temp}
Gas atmospheres in galaxy clusters are multi-temperature environments. Central regions are often prone to rapid radiative cooling, leading to the presence of cold gas phases mixed with the warm and hot components \citep[e.g.,][]{Fabian2003,Churazov2003,Sanders2016,Pinto2016,Hitomi2018}, while mergers and accretion of matter continuously supply outer cluster regions with multiphase gaseous structures \citep[e.g.,][]{Roncarelli2013,Angelinelli2021,Gouin2023}, driving shocks and turbulence. Additionally, 3D temperature gradients, including rapid temperature drop beyond $0.1-0.2 r_{200}$ \cite[e.g.,][]{Markevitch1998,Vikhlinin2006}, will be imprinted on the observed 2D temperature. All of these will lead to broad temperature distributions at a given radius and, if not taken into account in the spectral modelling, may bias the recovered gas characteristics, especially in the cluster outskirts. Similar biases on metal abundances and spectroscopic-like
temperatures have been discussed in the literature \citep[e.g.,][]{Buote2000,Rasia2008,Simionescu2009,Zuhone2023a}.

\begin{figure}
\centering
\includegraphics[width=0.9\linewidth]{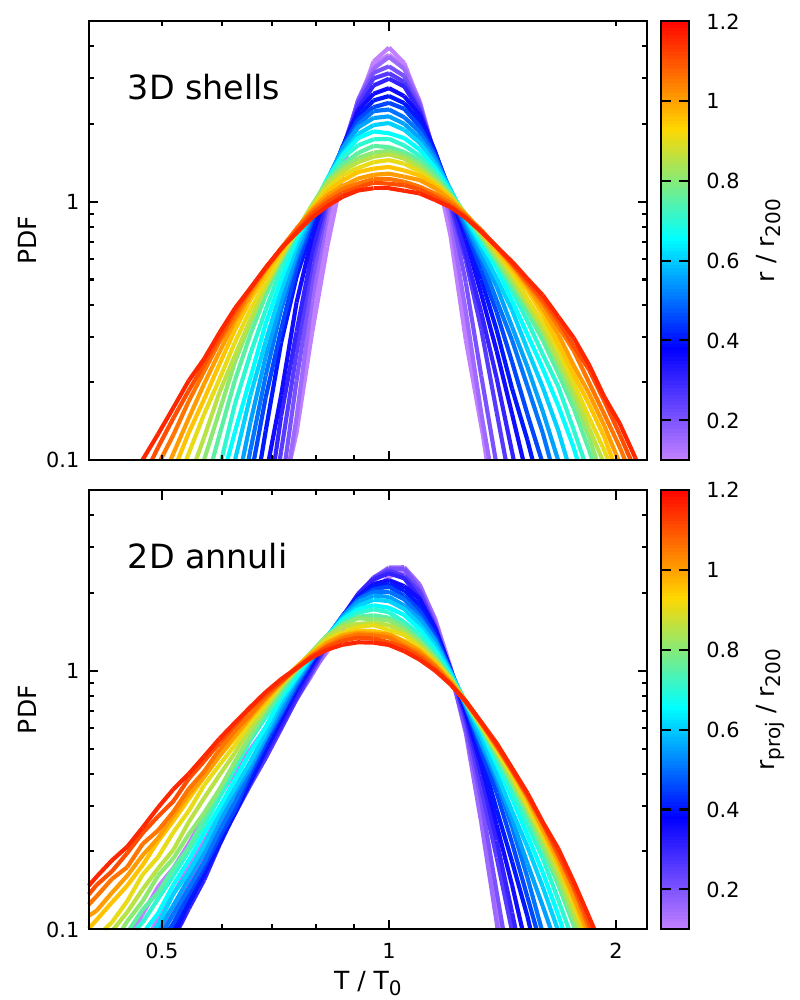}
\caption{Sample median PDFs of the ICM temperature in radial shells weighted by gas mass (top) and in 2D annuli along the LOS weighted by broad-band X-ray emissivity (bottom). The distance is shown in colour, $T_0$ is the median temperature in radial shells (mass-weighted) or annuli (X-ray emissivity weighted; see Section~\ref{sec:sample:temp}).}
\label{fig:temp_pdf}
\end{figure}

\begin{figure*}
\centering
\includegraphics[width=0.9\linewidth]{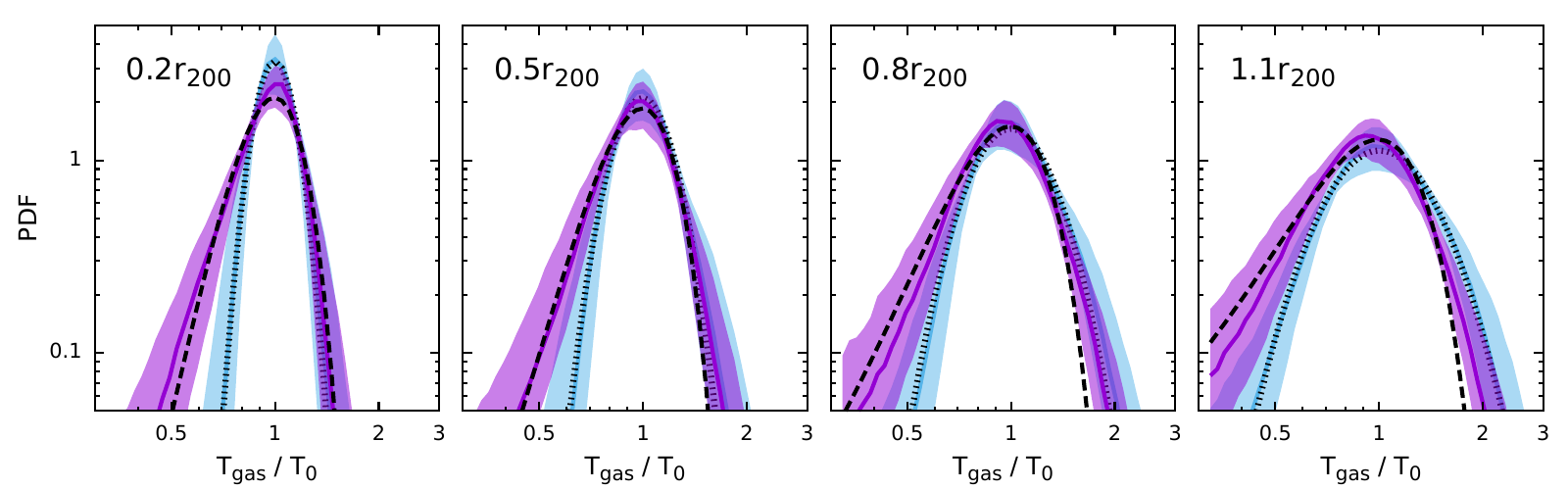}
\caption{PDF of temperature fluctuations at four different radii shown in Fig.~\ref{fig:temp_pdf}. Solid lines compare their median profiles in radial shells (blue) and in 2D annuli (purple). The corresponding bands indicate the 10th and 90th percentiles, illustrating the level of scatters over the cluster sample. The black dotted and dashed lines show our best-fit log-normal and Gaussian distributions for the profiles, respectively (see Section~\ref{sec:sample:temp}). }
\label{fig:temp_pdf2}
\end{figure*}

In Fig.~\ref{fig:temp_pdf}, we characterize ICM temperature fluctuations in our entire TNG300-cluster sample relative to the median temperature. The top panel shows sample median profiles of the probability distribution function (PDF) for temperature fluctuation in narrow radial shells with bin width $0.05r_{200}$ weighted by gas mass. The PDFs can be well described by log-normal distributions \citep{Kawahara2007,Zhuravleva2013,Khedekar2013}, being broader at larger cluster radii (see also Fig.~\ref{fig:temp_pdf2}). The PDF distribution can be understood as arising from the cumulative effects of multiple processes (e.g., turbulence, shocks, merging structures) driving temperature fluctuations. We approximate the radial dependence of the PDF's variance within the radial range $\le1.5r_{200}$ by

\be
\sigma_{\rm logt}(r)=\sigma_{0}(e^{\eta r/r_{200}}-1) + \sigma_{\rm logt}(0),
\ee
where three best-fit parameters are $\sigma_{0}=1.0$, $\sigma_{\rm logt}(0)=0.08$, and $\eta=0.2$, respectively. We note that different numerical codes and sub-grid models used in the simulations may affect these parameters, but the log-normal shape of the temperature distribution is unlikely changed \citep{Rasia2014}. Moreover, the PDF of an individual cluster is not smooth. There are variations/scatters caused by (resolved) individual substructures, e.g., clumps and shocks, especially in the cluster outskirts. In Fig.~\ref{fig:temp_pdf}, we show only median profiles, where most scatters have been cancelled. Our relaxed cluster-only sample gives almost the same results (not shown in the paper). These profiles are referred to as the intrinsic temperature distribution of the ICM. In Section~\ref{sec:mock:model}, we will demonstrate that it is crucial to take this intrinsic temperature distribution into account to prevent systematic bias while fitting ICM model parameters (e.g., metallicity).

In the bottom panel of Fig.~\ref{fig:temp_pdf}, we show the temperature PDF again but in 2D annuli of projected temperature along the LOS, weighted by broad-band X-ray emissivity. They are directly connected with X-ray spectra and show asymmetric profiles due to the projection effect. The low-temperature tail of the distribution is mostly contributed by the low-temperature gas at a larger radius in the LOS (compare the top and bottom panels).

A direct comparison between PDF in 3D and in projection is shown in Fig.~\ref{fig:temp_pdf2}. The latter is no longer log-normal but closer to a Gaussian distribution because it is a sum of many log-normal distributions along the LOS. The best-fit Gaussian variance ranges from $\simeq0.2\ (0.1r_{200})$ to $0.3$ ($r_{200}$; see also the blue line in Fig.~\ref{fig:tsigma_params}).
We have noticed that a skew-normal distribution fits the PDF in the projection better, especially at the high temperature end. However, as we discuss in Section~\ref{sec:results:gas}, the skewness can hardly be recovered when fitting mock X-ray spectra. Therefore, we no longer consider it in this work.

\section{Mock X-ray Images and Spectra} \label{sec:mock}

To illustrate quantitatively the unique opportunities of studying galaxy clusters with soft X-ray calorimeters, we mainly focus on the LEM mission concept. It provides a representative example of several perspective X-ray spectroscopy missions, taking advantage of the large grasp and high-spectral resolution, ideal for probing weak X-ray signals of cosmic structures. However, our main results are general and applicable to other similar missions (e.g., HUBS, Super\,DIOS; see \citealt{Cui2020,Yamada2018}).


\subsection{Mock X-ray observation} \label{sec:mock:mock}

We generate LEM mock cluster event files using \texttt{pyXSIM}\footnote{\href{https://hea-www.cfa.harvard.edu/~jzuhone/pyxsim}{https://hea-www.cfa.harvard.edu/\~{}jzuhone/pyxsim}} and \texttt{SOXS}\footnote{\href{https://hea-www.cfa.harvard.edu/soxs}{https://hea-www.cfa.harvard.edu/soxs}} Python packages, which simulate observations of X-ray sources with current and future X-ray telescopes. Specifically, \texttt{pyXSIM} samples synthetic photons according to gas distributions in TNG clusters with a Monte Carlo approach based on the PHOX
algorithm \citep{Biffi2012,Biffi2013}, assuming the APEC model. Resonant scattering, photoionization, and non-equilibrium ionization are ignored and likely subdominant compared to other uncertainties. Then, \texttt{SOXS} convolves the sky-projected event positions and energies with an instrumental response ($2\eV$ energy resolution unless stated otherwise). 
To properly capture projection effects, we include all gas cells with zero star formation rate, gas temperature $T_{\rm gas}>10^{5.5}\K$, and density $\rho_{\rm gas}<0.5\,{\rm cm^{-3}}$ (close to the star formation density threshold in the TNG simulations)\footnote{These conditions are used to exclude a small number of unphysically overdense, isolated, and X-ray bright gas cells in the simulation that would not be expected to exist in real clusters but are a numerical artifact. They represent a very small fraction of the mass and X-ray emission of the cluster. Performing such cuts is a common procedure in similar works \citep{Rasia2013,Barnes2021,Pop2022,Zuhone2023b}} within at least $4r_{200}$ of a cluster when generating the mock.

\begin{figure}
\centering
\includegraphics[width=1\linewidth]{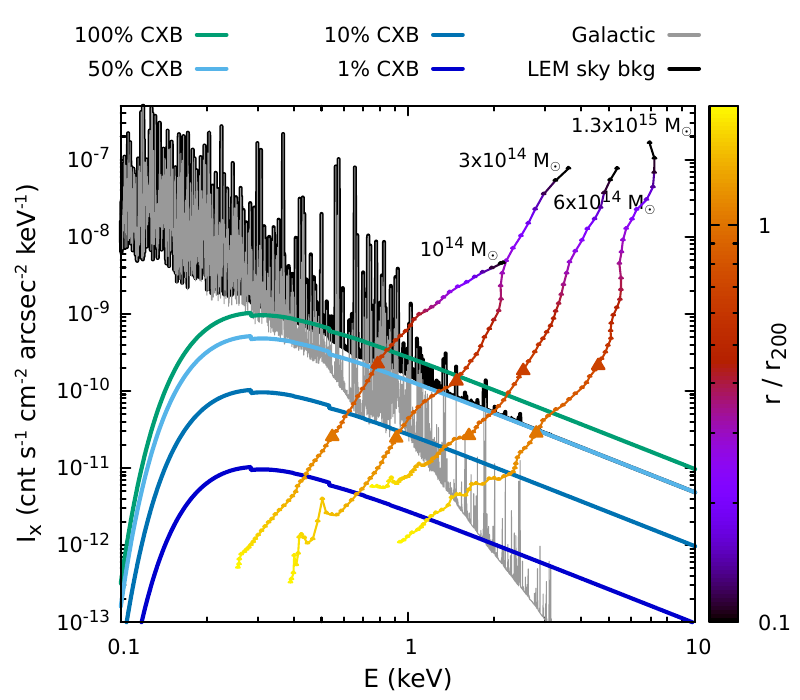}
\caption{Sky background model. The spectrum in grey indicates the Galactic foreground. Four solid curves show $100$, $50$, $10$, and $1$ per cent of the CXB described by an absorbed power law model. LEM sky background is adopted as the sum of unresolved CXB ($50$ per cent) and Galactic radiation (black spectrum). For comparison, broad-band X-ray surface brightness $I_{\rm x}(r)$ vs. X-ray temperature $T_{\rm x}(r)$ of three relaxed galaxy clusters listed in Table~\ref{tab:cluster_sample} and one of the most massive TNG300 clusters (ID\,11748) are overlaid as the colour curves with points. The colour encodes the cluster radius. The pairs of big triangles indicate $r_{500}$ and $r_{200}$, respectively. The ICM X-ray signal from the cluster outskirts is weak and has a soft spectrum (depending on the halo mass). The large grasp and high spectral resolution are vital to measuring their signals (see Section~\ref{sec:mock:mock}).
}
\label{fig:bkg}
\end{figure}

\begin{figure*}
\centering
\includegraphics[width=0.9\linewidth]{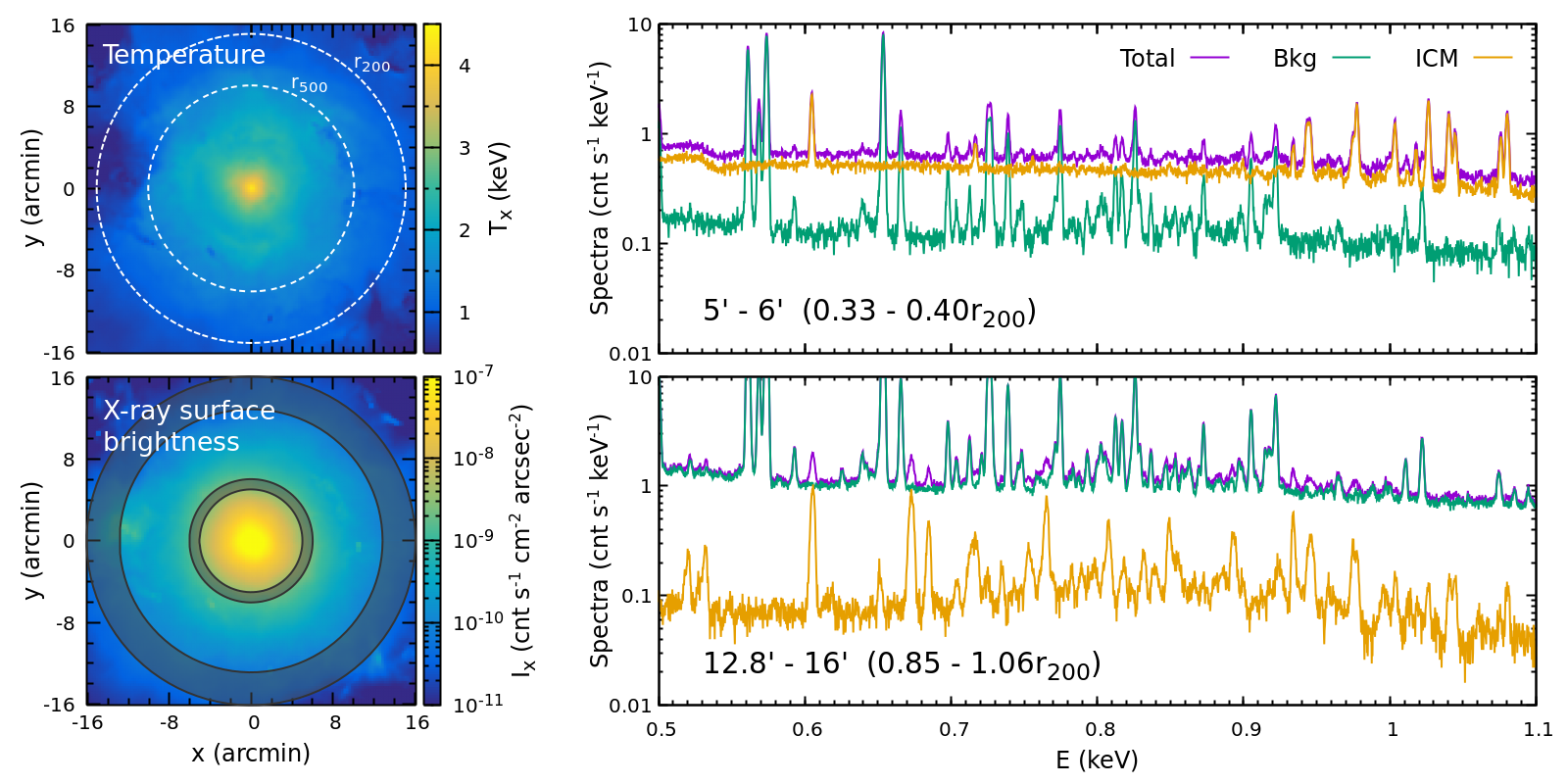}
\caption{Mock LEM spectra from a TNG300 cluster (CL-RM) at redshift $0.08$. The left two panels show distributions of X-ray-weighted temperature and X-ray surface brightness of the cluster. Two white circles indicate $r_{500}$ and $r_{200}$. The right panels show spectra extracted from two annuli marked in the bottom left panel (darker regions). The green, yellow, and purple curves represent the background (including both sky and instrumental contributions), ICM emission, and the total spectrum, respectively. This figure illustrates the key role of high spectral resolution in detecting a signal (i.e., several strongest individual lines) from ICM, allowing unambiguous separation of the cluster's and Milky Way's signals, even in low surface brightness, outer regions of clusters (see Section~\ref{sec:mock:mock}).}
\label{fig:spec_example}
\end{figure*}

Besides cluster emission, the background is also a major component of the X-ray signal. The background includes Galactic foreground, cosmic X-ray background (CXB), and detector background. Fig.~\ref{fig:bkg} shows spectra of the (1) Galactic foreground that includes both local hot bubble and gas halo \citep{McCammon2002} and (2) CXB approximated as an absorbed power-law model with energy index $\Gamma=1.47$ \citep{Hickox2006}. For LEM, we expect that $\sim50$ per cent of CXB will be resolved, i.e., $\simeq50-75$ brightest point sources per FOV, corresponding to a loss of $\simeq1$ per cent detector area while excising them. The rest of the (unresolved) CXB, along with the Galactic emission, will be the sky background of LEM (black solid curve), which will be taken into account when fitting the spectra.
For comparison, we show radial profiles of X-ray surface brightness ($0.5-2\keV$) of four individual relaxed clusters as a function of their X-ray-weighted temperature $T_{\rm X}$ (colour lines with points). Two large triangles mark $r_{500}$ and $r_{200}$, respectively. The curves are overlaid in the figure by assuming that $k_{\rm B}T_{\rm X}$ corresponds to the dominant X-ray photon energy in the emission (approximately peak position of the bremsstrahlung continuum), where $k_{B}$ is the Boltzmann constant. This figure illustrates that ICM X-ray emissions from the cluster outskirts ($r>r_{500}$) are generally soft and weak except for very massive halos (e.g., $\gtrsim10^{15}\msun$). Below $2\keV$, it is non-trivial to measure the continuum outside $r_{500}$. High-spectral resolution along with a large grasp are essential to measure the ICM there via the detection of individual emission lines. Finally, for the detector background, we simply assume a flat spectrum $1\,{\rm cnt\,s^{-1}\keV^{-1}\,FOV^{-1}}$, based on conservative scaling of the Athena X-IFU estimates \citep{Barret2020}, which corresponds to $\simeq10^{-10}\,{\rm cnt\,s^{-1}\,cm^{-2}\,arcsec^{-2}\keV^{-1}}$ around $0.5-1\keV$.

Examples of our mock LEM spectra are shown in Fig.~\ref{fig:spec_example}, extracted from two annuli marked in the bottom left panel. Within $r_{500}$, the cluster's signal (yellow line) dominates the spectrum. Both continuum and emission lines (e.g., O\,VIII and Fe-L complex) are well resolved. However, this is not the case outside $r_{500}$. Near $r_{200}$, the ICM continuum is more than 10 times weaker than the background (see the bottom right panel). Thanks to the cluster's proper redshift, multiple strong lines from the cluster (e.g., O\,VII, O\,VIII, and Fe\,XVII) are redshifted and separate unambiguously from the Milky Way's signals, providing an observational window to map the ICM in the cluster outskirts.

\subsection{Spatially identifying cold clumps} \label{sec:mock:mask}

\begin{figure}
\centering
\includegraphics[width=0.9\linewidth]{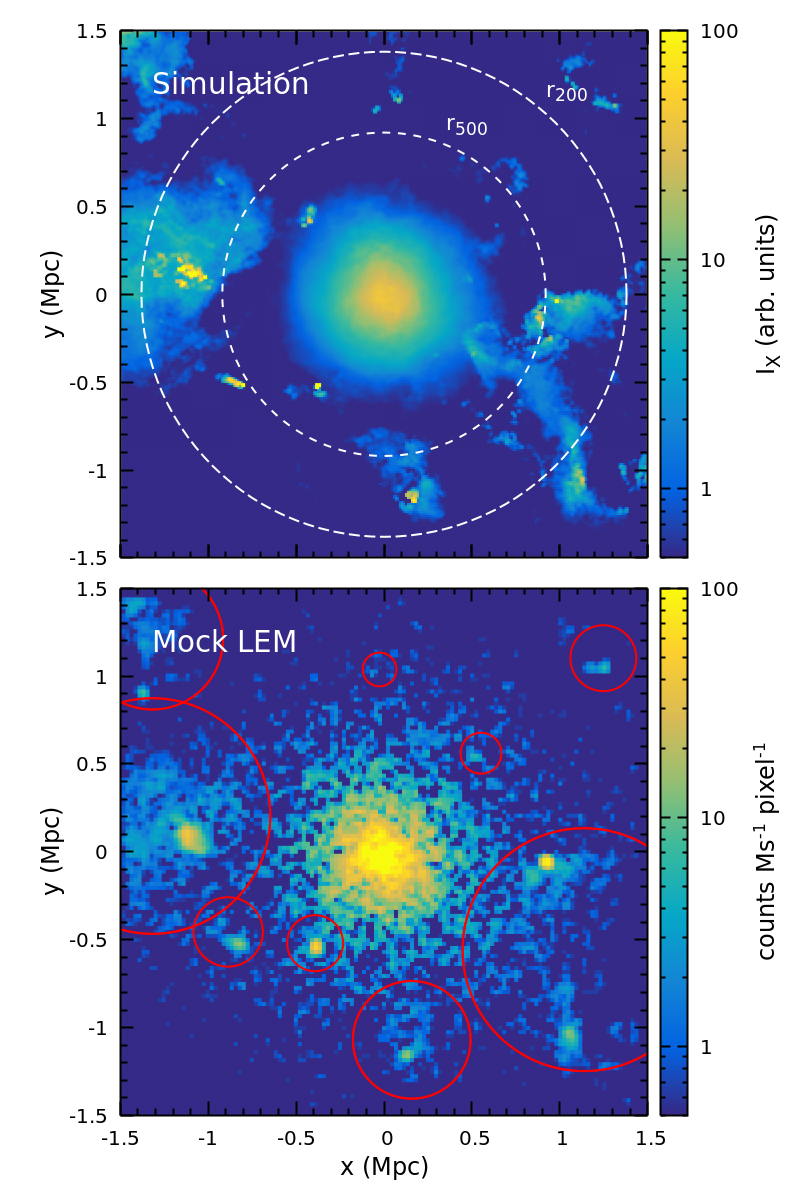}
\caption{\textit{Top panel:} modelled X-ray surface brightness from emission lines between $510$ and $540\eV$ (observer frame) of the cluster CL-RM at redshift $z=0.08$ (no continuum and instrumental response). In the cluster frame, this energy interval includes the O\,VII lines near $570\eV$. Two white circles indicate $r_{500}$ and $r_{200}$, respectively. Cold-gas clumps from stripped infalling galaxies and/or groups are clearly seen in this narrow-band image due to their strong O\,VII emissions. However, the central extended bright halo is largely contributed by Fe\,XXIV line emissions from the hot ICM. \textit{Bottom panel:} $2\Ms$ mock LEM image -- a difference between two narrow energy bands to remove the continuum (see Section~\ref{sec:mock:mask} for more details). It reproduces well the simulation shown in the top panel, demonstrating the robustness of identifying cold clumps with a LEM narrow-band image. The red circles mark visually-identified clumps, which would be masked when extracting spectra as one of our strategies in the study (see Section~\ref{sec:mock:model}). }
\label{fig:map_o7}
\end{figure}

\begin{figure}
\centering
\includegraphics[width=0.9\linewidth]{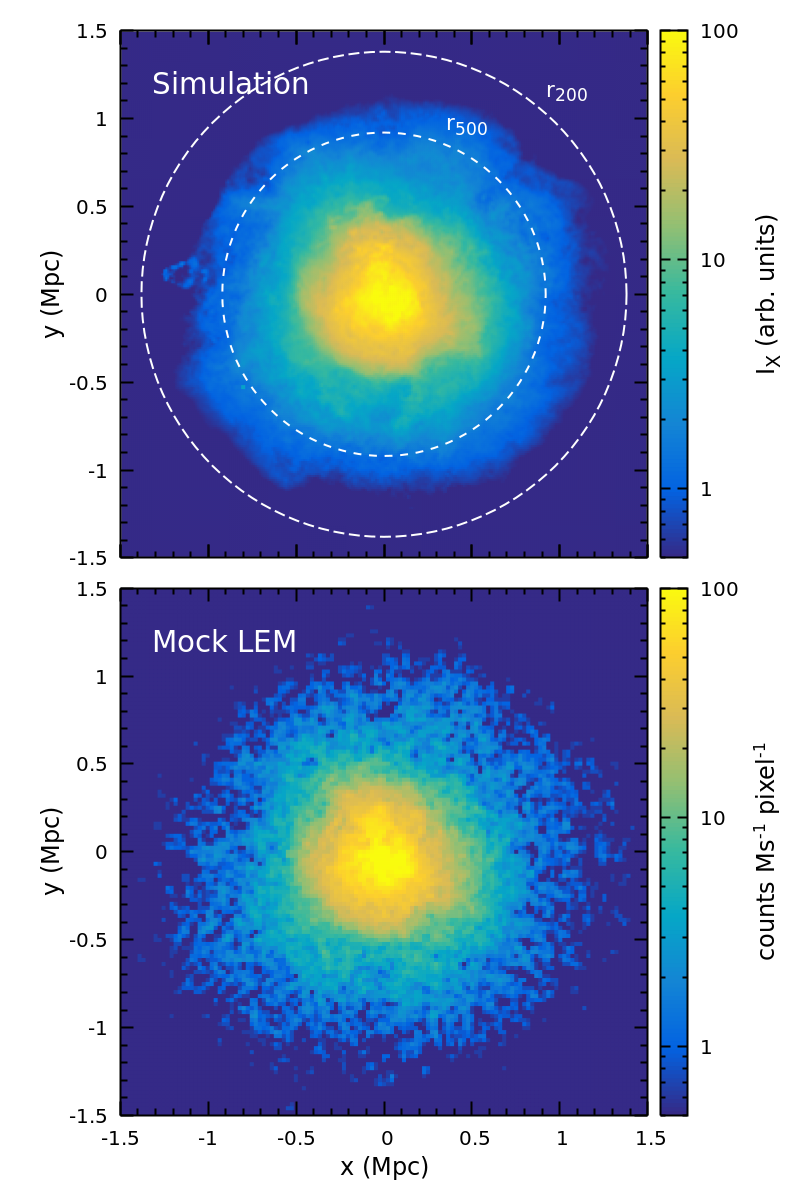}
\caption{Similar to Fig.~\ref{fig:map_o7} but for the line Fe\,XXIII near $975\eV$, tracing the hot ICM of the cluster (see Section~\ref{sec:mock:mask}). }
\label{fig:map_fe23}
\end{figure}

The ICM is multiphase, especially in the cluster outskirts, as discussed in Sections~\ref{sec:sample:vsigma} and \ref{sec:sample:temp}. It is possible to trace different gas phases through narrow-band X-ray imaging (see Fig.~\ref{fig:apec_spec}). For instance, relatively cold and dense clumps could be seen through the mapping of O\,VII triplet lines near $570\eV$ that are largely populated by $T_{\rm gas}\simeq0.1-0.5\keV$ gas, Fe\,XVII ($826\eV$) by $0.3-1\keV$ gas, and Fe\,XXIII ($1050\eV$) by $\gtrsim1\keV$ gas.

Fig.~\ref{fig:map_o7} shows narrow-band images of the CL-RM cluster around the O\,VII line. The top panel shows a simulated prediction of the surface brightness from emission lines only between $510$ and $540\eV$ in the rest frame of the observer. In the cluster frame, this band contains O\,VII lines. The continuum and instrumental response are not included on the map. Many cold gas clumps are clearly seen through their strong O\,VII emissions. Most of them are stripped material from infalling galaxies and/or groups. The biggest substructure within the FOV is in the east of the cluster, visible even in the broad-band X-ray and temperature distributions (see Figs.~\labelcref{fig:spec_example,fig:hist_vr}). Note that the extended bright halo in the central region is not from O\,VII but weak Fe\,XXIV lines emitted from the hot ICM ($T_{\rm gas}\gtrsim1\keV$).

For comparison, the bottom panel of Fig.~\ref{fig:map_o7} shows a $2\Ms$ mock LEM surface brightness $2I_{510-540}-I_{495-555}$, where $I_{510-540}$ ($I_{495-555}$) is the surface brightness of the ICM in the energy band $510-540\eV$ (495-555\eV). Subtraction of these two narrow bands removes the continuum. For simplicity, we did not include any background in the image since the background is assumed to be uniform spatially in our model, and there are no strong Galactic emission lines in the adopted energy bands. The background will only contribute Poisson noise to our results. The mock image captures all the major features of the simulated image, demonstrating the robustness of applying LEM narrow-band images to resolve cold clumps in the ICM up to $r_{200}$ and even beyond. The red circles in the figure mark the visually-identified clumps, which will be masked when extracting the spectra as one of our strategies in the study (see Section~\ref{sec:mock:model}). The mask used here is rather aggressive -- $\simeq1/3-1/2$ of the region outside $r_{500}$ is excluded.

Fig.~\ref{fig:map_fe23} shows similar images but for Fe\,XXIII over a narrow energy band $972-982\eV$ (i.e., $2I_{972-982}-I_{967-987}$), revealing distribution of the hot gas phase of the cluster up to $r_{200}$, distinguishable from the cold clumps' distribution in Fig.~\ref{fig:map_o7}.

\subsection{Spectral analysis and Xspec model} \label{sec:mock:model}

We extract ICM X-ray spectra from mock event files and generate the background component by using the fakeit command in the X-ray spectral fitting package \texttt{Xspec} \citep{Arnaud1996} while including the Galactic foreground, 50 per cent CXB (as 50 per cent is removed as discrete, detected sources), and detector background as described in Section~\ref{sec:mock:mock} (see examples in Fig.~\ref{fig:spec_example}).

We fit the total spectra to a combined model including both cluster and background components within the energy range $0.4-1.5\keV$, assuming Cash statistics \citep{Cash1979}. Fitting a wider energy band does not provide much additional information since there are no strong emission lines. Fitting errors are quoted at the $90$ per cent confidence level unless otherwise noted. For simplicity, we fix all the parameters involved in the background so that the background contributes merely statistical uncertainties to the fitting results. This is a reasonable assumption for the purpose of this study, given that the majority of strong lines from the cluster have been redshifted away from the Galactic foreground lines. In reality, however, temporal and spatial variations of the background could be taken into account by treating some (if not all) background parameters free. It is expected that our knowledge of the Milky Way foreground will soon be advanced through the ongoing SRG/eROSITA and \textit{XRISM} missions \citep{Predehl2021,XRISM2020}.

We have shown in Section~\ref{sec:sample:temp} that the distribution of the X-ray-emissivity-weighted ICM temperature along the LOS can be approximated as a Gaussian distribution. To account for this, we construct an \texttt{Xspec} model \texttt{gaussbvapec}. It is similar to the built-in model \texttt{vgadem} with only two differences: (1) velocity broadening is included as in the \texttt{bvapec} model and (2) standard deviation of the Gaussian temperature distribution normalized by the mean temperature is a free parameter ($\sigma_{\rm temp}$) of the model. In Section~\ref{sec:results:metal}, we demonstrate that such an extended temperature distribution plays an important role in suppressing bias in metal abundances based on the single-temperature \texttt{bvapec} model.

Besides the intrinsic temperature spread of the ICM, there are also resolved cold gas clumps that can significantly impact our fitting results. We apply two strategies to deal with them and compare their pros and cons:
\begin{itemize}
  \item We have demonstrated in Section~\ref{sec:mock:mask} that one can apply a mask to exclude the majority of the cold clump contributions from the extracted spectra (see, e.g., Fig.~\ref{fig:map_o7}). The remaining (unresolved) cold gas makes a minimal contribution to the spectra. We then need only a hot gas component in the fitting model, either \texttt{gaussbvapec} (\texttt{1g} for short hereafter) or \texttt{bvapec} (\texttt{1b}), to fit the ICM spectrum. The latter is, however, used only for comparison.
  \item High-resolution X-ray spectroscopy provides the capability of resolving cold clumps spectroscopically. It requires a major component and $N$ additional colder components to fit the hot ICM and cold clumps, respectively. We apply a \texttt{gaussbvapec} model for the former and place a prior constraint on its mean temperature parameter (i.e., $0.5-10\keV$). We start our fitting from $N=0$. In the following iterations, we carry out a new fitting each time by adding one more clump component and check if it makes a non-negligible contribution to the cluster signals, i.e., its normalization is larger than 1 per cent of the major component. This way is more robust than checking only $\chi^2$, since background always dominates the spectrum in the cluster outskirts. For simplicity, we model each clump component by only a \texttt{bapec} model with a prior temperature constraint as $0.05-1\keV$. We stop the iteration and take the previous best-fit results if the newly added cold component is insignificant (weaker than 1 per cent of the major component). In our experiment, $N\leq2$ always provides sufficiently good fittings for our sample. We refer to this model as \texttt{1g2b} in the following discussions (see Table~\ref{tab:fitting_model} for a summary of the models used in the study).
\end{itemize}
In general, we find that (1) with a mask, we are able to constrain metal abundances of the hot ICM better (see Section~\ref{sec:results:metal}); (2) without a mask, the \texttt{1g2b} model accurately captures the multiphase gas in galaxy clusters, providing a better constraint on the velocity dispersion especially in the cluster outer regions since more photons are used in the fitting (see Section~\ref{sec:results:vsigma}). The spectroscopically resolved cold clumps provide important phase-space information of the infalling substructures in galaxy clusters (see Section~\ref{sec:results:clumps}).

\begin{table}
\centering
\begin{minipage}{\linewidth}
\centering
\caption{A summary of our \texttt{Xspec} models used to fit the LEM mock spectra (see Section~\ref{sec:mock:model}). }
\label{tab:fitting_model}
\begin{tabular}{lcccc}
  \hline
  \multirow{2}{*}{\makecell{Model \\ name}}  &
  \multirow{2}{*}{\makecell{With \\ mask}}  &
  Hot comp.   &
  \multicolumn{2}{c}{Additional cold comp.}
\\
    & &
    (1st) &
    (2nd &
    3rd)
\\ \hline
 1g   & Yes & \texttt{gaussbvapec}  &  -  &  -  \\
 1b   & Yes & \texttt{bvapec}       &  -  &  -  \\
 1g1b\footnote{This model is only used in Section~\ref{sec:results:filaments}.}
      & No  &\texttt{gaussbvapec}  &  \texttt{bapec}  & -  \\
 1g2b & No  &\texttt{gaussbvapec}  &  \texttt{bapec}  & \texttt{bapec}  \\
\hline
\vspace{-18pt}
\end{tabular}
\end{minipage}
\end{table}
%


\section{Science Drivers} \label{sec:results}

In this section, we discuss six key science drivers for future LEM-type X-ray missions on the ICM science and highlight the capabilities of high-resolution X-ray spectroscopy in pushing ICM measurements and understanding their underlying physics to a new level.

Based on our TNG300 cluster sample, we present (1) measurements of gas thermodynamic (Section~\ref{sec:results:gas}), metal abundance (Section~\ref{sec:results:metal}), and non-thermal pressure (Section~\ref{sec:results:vsigma}) profiles up to $r_{200}$, (2) reconstruction of gas phase-space distributions (Section~\ref{sec:results:clumps}), and (3) mapping of gas LOS velocity distributions to constrain ICM turbulence properties (Section~\ref{sec:results:vlos}) and detect ICM-filament interactions (Section~\ref{sec:results:filaments}). We provide strategies to optimize data analysis and interpretation. Unless otherwise stated, we assume that the observation duration of each cluster is $t_{\rm exp}=2\Ms$. X-ray photons are radially binned to measure the ICM radial profiles. The minimum number of counts from the cluster in each bin and the minimum radial size of the bin are $10^5\Ms^{-1}$ and $1'$, respectively.

\subsection{Gas thermodynamics of the ICM} \label{sec:results:gas}

\begin{figure}
\centering
\includegraphics[width=0.9\linewidth]{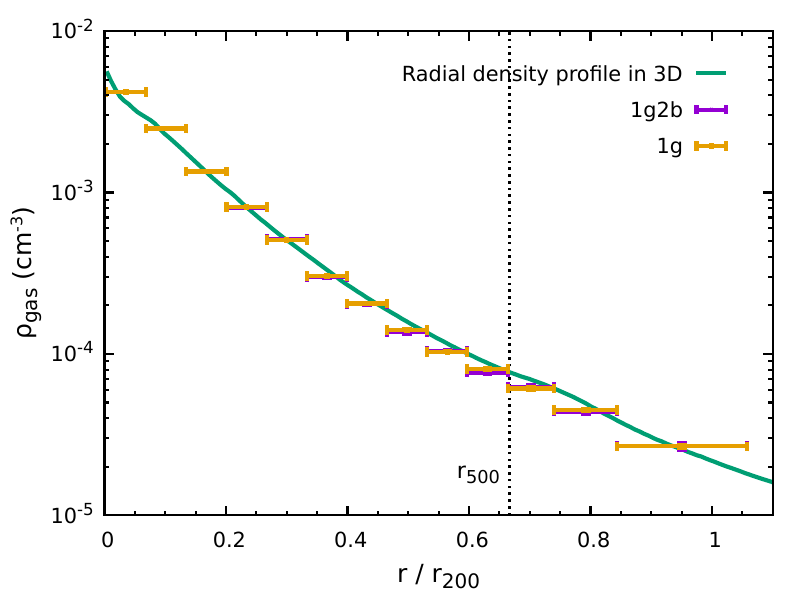}
\caption{A comparison of the LEM best-fit deprojected density radial profile of the hot ICM (points with error bars) with the simulation prediction (green line) for the cluster CL-RM. The model includes only gas cells with $T_{\rm gas}\ge1\keV$ to exclude contributions from cold clumps. The vertical dashed line marks $r_{500}$. The LEM measurements accurately recover the simulation, independent of the strategies for resolving cold clumps (see Section~\ref{sec:results:gas}). }
\label{fig:gas_profile}
\end{figure}

Fig.~\ref{fig:gas_profile} shows the LEM best-fit gas density radial profile of the cluster CL-RM and compares it with the true profile in simulation (green line). The deprojected radial profiles are reconstructed from the best-fit normalization of the hot ICM component using the standard onion-peeling approach \citep{Kriss1983}. The statistical uncertainties are small ($\lesssim5$ per cent near $r_{200}$) and barely visible in the plot. The simulation curve is estimated by excluding all gas cells colder than $1\keV$ to remove cold clumps that create strong spikes on top of the global radial profile. The LEM measurements recover the density radial profile of the ICM. Different fitting approaches show almost identical results. Determining slopes of gas density profiles outside $r_{500}$ is vital to understanding the evolution of the runaway merger shocks and how they re-shape matter distributions in the cluster outskirts \citep{Zhang2019,Zhang2020}. Cosmological simulations suggest a steep gas density profile near the virial radius ($\propto r^{-3}$ or steeper; see, e.g., \citealt{Roncarelli2006,Lau2015}), implying the existence of a ``habitable zone'' for runaway shocks that can power radio relics in the peripheries of the cluster \citep[e.g.,][]{Lyskova2019}. However, current X-ray measurements are often contaminated by the complexity of clumpiness \citep[e.g.,][]{Nagai2011}. In our experiment, we do not see any significant gas density overestimation caused by such an issue, thanks to LEM's capability of separating the hot ICM and cold clumps spectroscopically. Fig.~\ref{fig:gas_profile} shows that high-resolution X-ray spectroscopy accurately recovers the normalization and slope of the ICM density radial profile up to $r_{200}$ and beyond, which will allow distinguishing different slopes of gas density profiles along filamentary and non-filamentary directions (see, e.g., \citealt{Vurm2023} and also Section~\ref{sec:results:filaments}).

\begin{figure}
\centering
\includegraphics[width=0.9\linewidth]{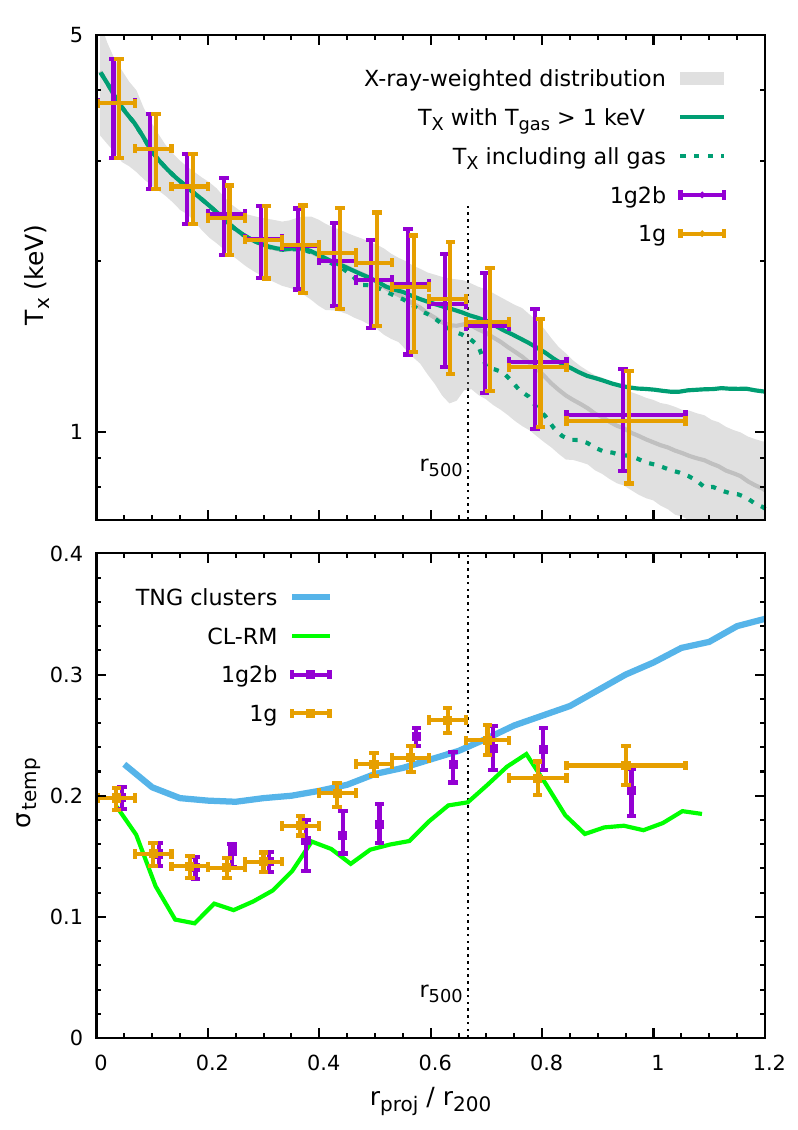}
\caption{\textit{Top panel}: LEM measurements of the projected temperature profile of the ICM in CL-RM (points with error bars). The vertical error bars show the intrinsic temperature spread of the ICM $T_{\rm gas}\times\sigma_{\rm temp}$ rather than the statistical uncertainties, which are very small ($\sim$ a few per cent). The grey band shows the simulated broad-band X-ray-weighted temperature distribution. The upper and lower boundaries indicate $\pm1\sigma$ of the distribution, and the middle line shows the mean. Cold clumps are excluded using the same mask as in the mock LEM analysis. The green lines show the modelled X-ray-weighted temperature profiles without any mask, while the solid line includes only gas cells with $T_{\rm gas}>1\keV$. \textit{Bottom panel}: Best-fit standard deviation $\sigma_{\rm temp}$ of the ICM temperature distribution normalized by the mean gas temperature in the \texttt{gaussbvapec} model in each annulus. The green and blue lines show the predicted profile of CL-RM and the average over the entire TNG300-cluster sample, respectively. This figure shows that, in addition to the mean temperature, LEM can recover $\sigma_{\rm temp}$, providing essential information about the ICM temperature structure (see Section~\ref{sec:results:gas}). }
\label{fig:tsigma_params}
\end{figure}

In Fig.~\ref{fig:tsigma_params}, we show the best-fit ICM projected temperature profile (the top panel). The error bars represent the intrinsic temperature spread of the ICM $T_{\rm gas}\times\sigma_{\rm temp}$ rather than the uncertainty of the temperature statistics (smaller than $\simeq3$ per cent at $r_{200}$). For comparison, we estimate the broad-band X-ray-weighted temperature histogram as a function of the projected radius shown as the grey band. We use the same mask as in the mock LEM analysis to exclude clumps (see Fig.~\ref{fig:map_o7}). The upper and lower boundaries indicate $\pm1\sigma$ of the distribution, and the middle line shows the mean temperature. Our mock measurements agree with the underlying model. They slightly overestimate the temperature near $r_{200}$ by $\lesssim10$ per cent due to the complex temperature structure in the cluster outskirts (e.g., deviation from Gaussian; see Fig.~\ref{fig:em_hist_180645}). Our \texttt{1g} and \texttt{1g2b} models show almost identical results.

The top panel of Fig.~\ref{fig:tsigma_params} shows a direct view of the intrinsic temperature distribution of the ICM. It cannot be characterized merely by a single temperature at a given radius. The green lines show the modelled broad-band X-ray-weighted temperature profiles without any mask. In particular, the solid line includes only the gas cells hotter than $1\keV$. The lines show up to $\sim40$ per cent deviations around $r_{200}$ from each other (also compared to the grey line). It is non-trivial to unambiguously define the X-ray temperature of the ICM in the cluster outskirts, even in original simulations; it is sensitive to the chosen criteria of separating the hot ICM and cold gas components. For this reason, a two-parameter model (i.e., temperature peak and spread) provides a more comprehensive characterization of the ICM temperature structure.

In the bottom panel of Fig.~\ref{fig:tsigma_params}, we show a direct comparison of our best-fit $\sigma_{\rm temp}$ with the simulation. The green and blue lines show a modelled profile of the cluster CL-RM and an average over our TNG300 cluster sample (see Fig.~\ref{fig:temp_pdf2}). The recovered $\sigma_{\rm temp}$ is encouragingly close to the prediction and is not sensitive to the scheme used to exclude the resolved cold clumps, given the complexity of resolving the ICM temperature structure. The well-relaxed CL-RM shows a slightly smaller $\sigma_{\rm temp}$ than the average but a similar radial profile shape. This result demonstrates high-resolution spectroscopy's capability of measuring the intrinsic gas temperature distribution, which could be used to explore turbulence, mixing, and gas transport properties of the ICM. Besides, the mild rise of the $\sigma_{\rm temp}$ profile in the innermost region of the cluster ($\lesssim0.2r_{200}$) likely reflects the active galactic nucleus (AGN) activity, which can be used to constrain the impact of feedback on the ICM.

\subsection{Metal abundances} \label{sec:results:metal}

When and how the hot and/or warm gas phase of the intergalactic medium (including the ICM) was enriched with metals is an important open question in large-scale structure evolution. Metal-abundance radial profiles in cluster outskirts provide a clean diagnostic for different enrichment scenarios for several reasons: (1) the cluster outer regions ($\gtrsim r_{500}$) are not affected strongly by the activity of the cluster's central supermassive black hole; (2) metal distributions in the cluster outskirts reflect the enrichment of gas being accreted recently from the cosmic web; and (3) early AGN feedback ($z>2-3$) that stirs and mixes metals outside of galaxies considerably impacts the slopes of abundance profiles outside $\sim0.5r_{200}$ (see, e.g., \citealt{Biffi2018} and also \citealt{Mernier2018} for a review).

\begin{figure}
\centering
\includegraphics[width=0.9\linewidth]{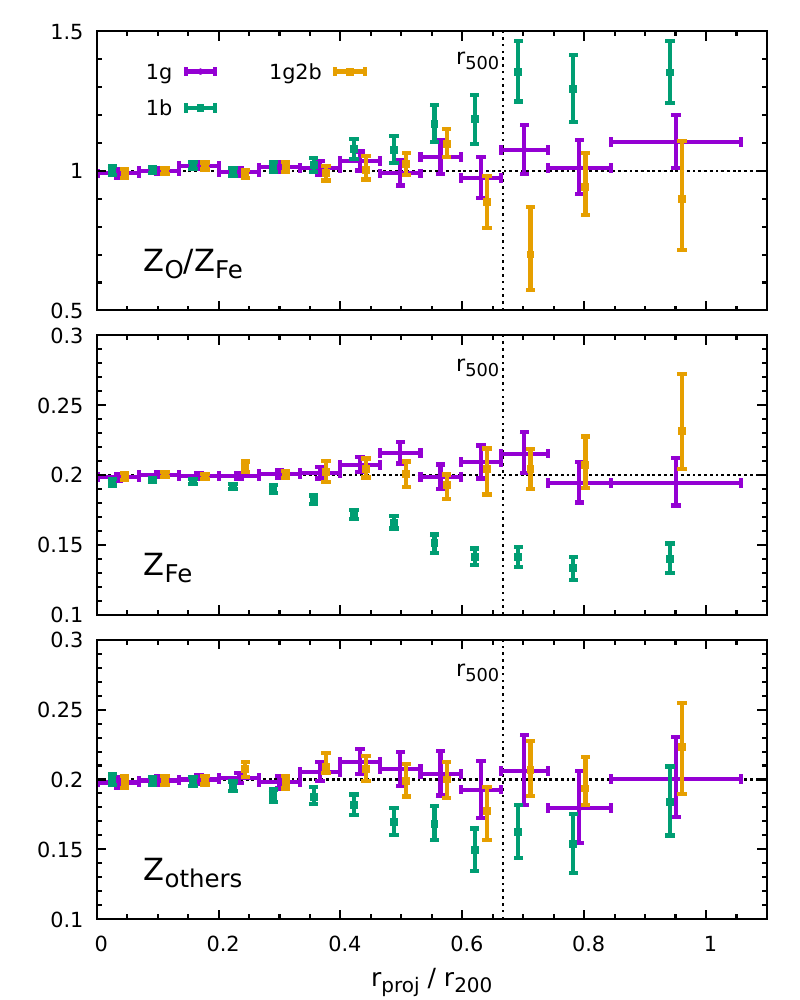}
\caption{Best-fit metal abundances of the cluster CL-RM while assuming a uniform metallicity distribution ($Z=0.2$). The panels show the abundance parameters fitted by our model: oxygen-to-iron ratio, iron abundance, and abundance of other elements (from the top to bottom). The horizontal dotted lines mark the input value. The \texttt{1g} model recovers the parameters the best, demonstrating that modelling a temperature distribution (e.g., a Gaussian distribution) is vital for correct measurements of abundance profiles (see Section~\ref{sec:results:metal}). }
\label{fig:metal_params}
\end{figure}

\textit{Suzaku} measured the radial profiles of Fe abundance in massive clusters (e.g., Perseus) up to $r_{200}$ based on Fe-K lines \citep[e.g.,][]{Werner2013,Urban2017}, suggesting a flat iron distribution extending from $\simeq0.3r_{200}$ to at least $r_{200}$ ($\simeq0.3$ solar). However, it is still under debate whether such a feature is universal from massive clusters to galaxy groups and if other, lighter elements (e.g., $\alpha$-elements) follow the same pattern \citep[e.g.,][]{Simionescu2015,Sarkar2022}. Future high-energy resolution X-ray missions will have the capability to address these questions, as discussed below.
In this study, we focus on recovering iron and oxygen abundances that emit the strongest lines in soft X-rays with high-resolution X-ray spectroscopy. While fitting the spectra, we use three parameters in the model: $Z_{\rm Fe}$, $Z_{\rm O}/Z_{\rm Fe}$, and $Z_{\rm others}$, representing iron abundance, oxygen-to-iron ratio, and the abundance of other elements, all with respect to solar.

Fig.~\ref{fig:metal_params} shows the results based on our baseline cluster model - with uniform metallicity ($Z=0.2$) throughout the cluster, where we can unambiguously characterize the biases and uncertainties in the measurements. We compare our best-fits from different fitting schemes. The horizontal dotted lines mark the input abundance value. The \texttt{1g} model correctly recovers all parameters up to $r_{200}$. In contrast, the \texttt{1b} model reveals prominent biases outside $0.3r_{200}$, $Z_{\rm Fe}$ in particular. The deviation of $Z_{\rm O}/Z_{\rm Fe}$ from unity is largely caused by the bias in Fe abundance. Masks are applied in both cases to spatially exclude the cold clumps' contribution (see Fig.~\ref{fig:map_o7}). The strong dependence of the Fe-L complex on gas temperature (see Fig.~\ref{fig:apec_spec} and also \citealt{Gastaldello2021}) largely explains why the single-temperature assumption leads to large biases in our fitting. It highlights the importance of modelling the ICM temperature structure (a Gaussian distribution) to accurately measure metal distributions. In Fig.~\ref{fig:metal_params}, the \texttt{1g} and \texttt{1g2b} models show similar constraints on $Z_{\rm Fe}$ and $Z_{\rm others}$. The latter reveals a larger scatter in $Z_{\rm O}/Z_{\rm Fe}$, because cold clumps generate both O\,VII and O\,VIII, complicating the spectral modelling and interpretation in the \texttt{1g2b} model. 

\begin{figure}
\centering
\includegraphics[width=0.9\linewidth]{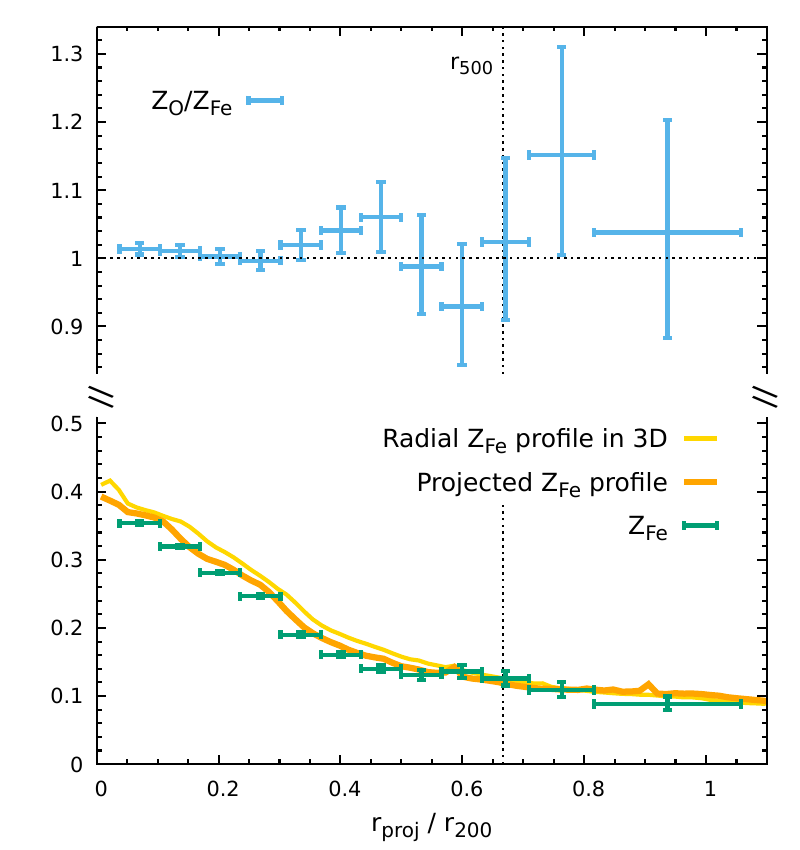}
\caption{Similar to Fig.~\ref{fig:metal_params} but assuming a non-uniform cluster metallicity directly from the cosmological simulation. The yellow and orange lines show the simulated Fe-abundance radial profile in 3D and the broad-band X-ray-weighted profile in projection where only gas cells with $T_{\rm gas}>1\keV$ are included. The points with error bars show our best-fit parameters $Z_{\rm Fe}$ (green) and $Z_{\rm O}/Z_{\rm Fe}$ (blue) from the \texttt{1g} model, fully consistent with the simulation input. The non-uniform metallicity distribution does not make the abundance recovery more difficult (see Section~\ref{sec:results:metal}). }
\label{fig:metal_tng_params}
\end{figure}

In Fig.~\ref{fig:metal_tng_params}, we explore a more realistic situation, where the metallicity varies spatially (i.e., from the simulation directly). It is known that the ICM metallicity in TNG300-1 does not converge to those in TNG100-1 due to the dependence of stellar mass on numerical resolution \citep[][]{Vogelsberger2018}. We include a factor of $1.6$ to adjust TNG300-1 to TNG100-1 as suggested by \citet{Vogelsberger2018} and still assume solar abundance ratios for all the elements, e.g., $Z_{\rm Fe}/Z=1$ \citep{Simionescu2015,Mernier2017}. The yellow and orange lines show the predicted Fe-abundance radial profile in 3D and projection (including only $T_{\rm gas}>1\keV$ gas and weighted by broad-band X-ray emissivity). The projection effect shows only a mild effect on the curve. The green and blue points show our best-fit $Z_{\rm Fe}$ and $Z_{\rm O}/Z_{\rm Fe}$ based on the \texttt{1g} model, fully consistent with the predictions, similar to those shown in Fig.~\ref{fig:metal_params}, although with larger error bars due to lower metallicity in this model (e.g., $Z\simeq0.1$ near $r_{200}$).

\subsection{Multiphase gas in the outskirts} \label{sec:results:clumps}

Although cold clumps complicate the X-ray spectra and recovery of the ICM temperature and metal abundances, they provide valuable information on the assembly history of galaxy clusters and star formation quenching (e.g., merger rate, gas stripping) and transport properties of the ICM (e.g., viscosity, thermal conduction). High-resolution spectroscopy enables cold clumps to be separated from the hot ICM spatially and spectroscopically, allowing direct measurements of the clump velocities and other physical properties.

\begin{figure}
\centering
\includegraphics[width=0.9\linewidth]{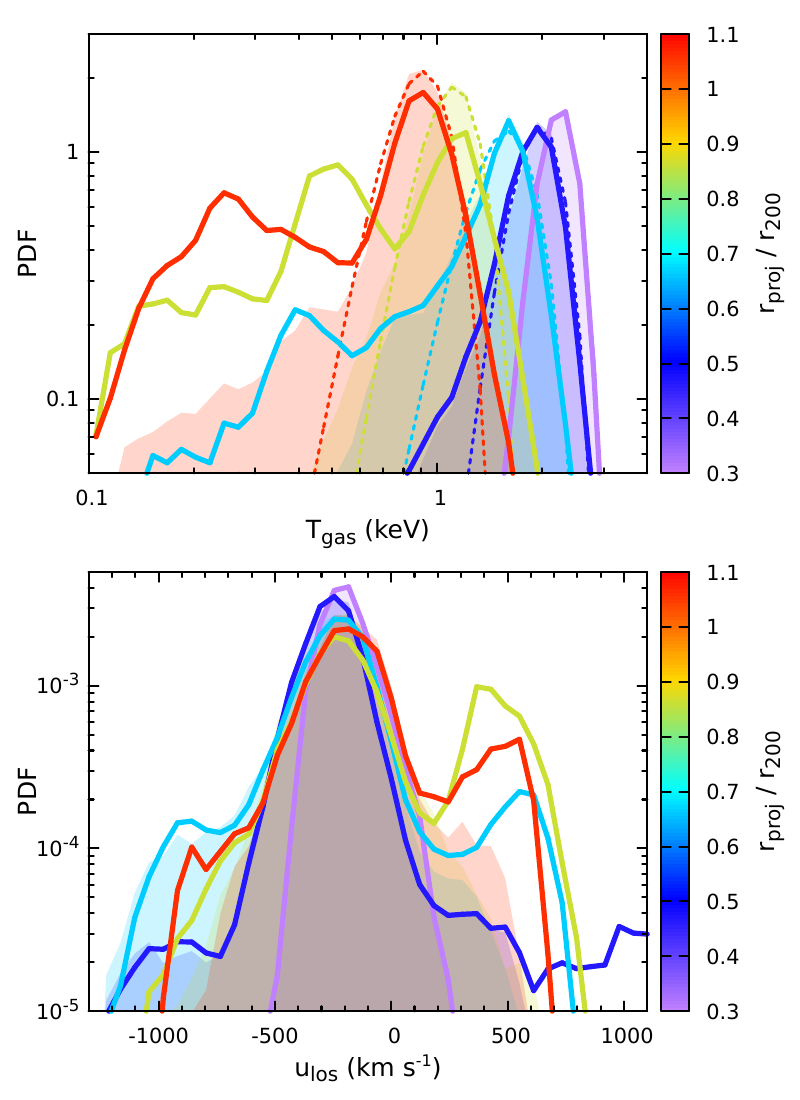}
\caption{Broad-band X-ray-weighted gas temperature PDF (top) and LOS velocity PDF (bottom) of the cluster CL-RM in projection. The line colour indicates the annular radius. The corresponding shaded regions show the PDFs after excluding clumps using the same mask as in the mock LEM analysis. The thin dotted lines in the top panel show the best-fit Gaussian distributions for the shaded temperature PDFs. This figure illustrates the complexity of gas structures in an individual cluster that encodes valuable information in the high-resolution X-ray spectra (see Section~\ref{sec:results:clumps}). }
\label{fig:em_hist_180645}
\end{figure}

\begin{figure}
\centering
\includegraphics[width=0.9\linewidth]{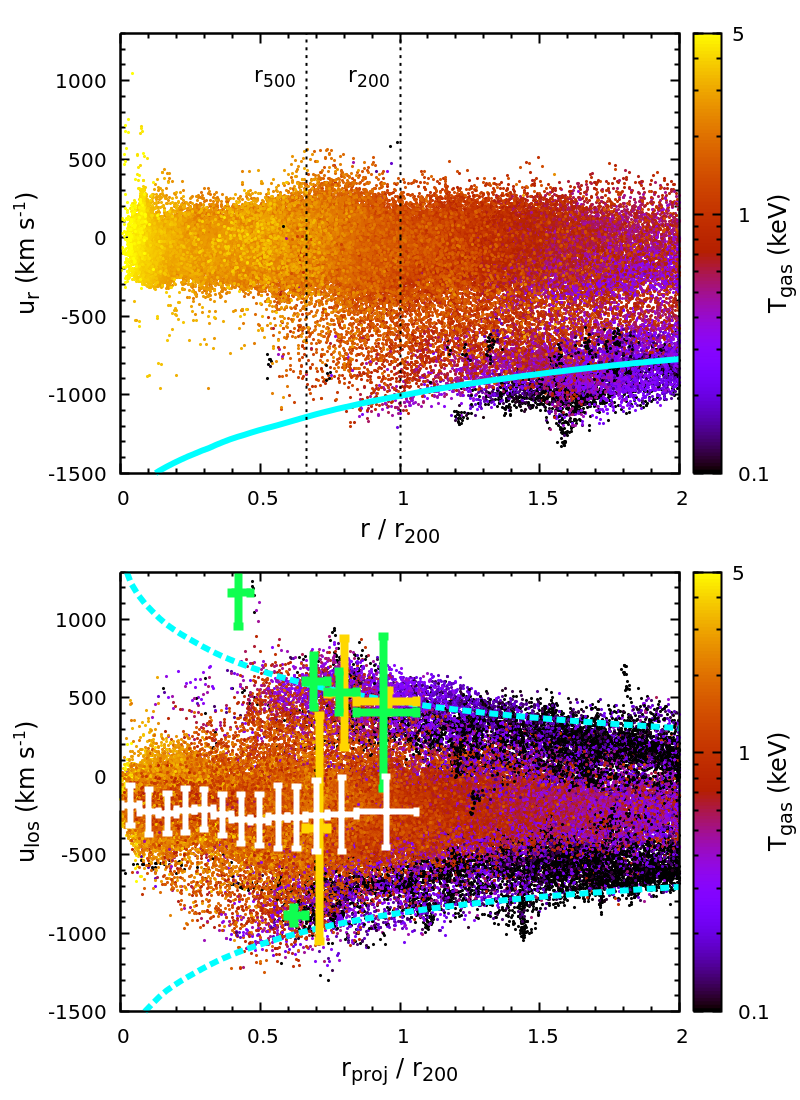}
\caption{\textit{Top panel}: phase-space distribution of gas cells in CL-RM -- radial position vs. radial velocity. The gas cells are randomly selected (1 out of 50) from the simulation, whose colour encodes the temperature. The cyan line shows the (inverse) escape velocity profile. \textit{Bottom panel}: phase-space distribution in projection, i.e., projected radius vs. LOS velocity, using the same set of gas cells. The cyan dashed lines show the maximum escape velocity projected along the LOS (see more details in Section~\ref{sec:results:clumps}). Our mock LEM measurements are overlaid as the points with error bars. The full bar sizes (vertical) reflect the FWHM of the velocity fields ($\propto\sigma_{\rm los}$). The white points show the primary component, the hot ICM. The green and yellow points show the second and third cold gas components. We slightly shift the green and yellow points horizontally for a clear view. The points overlap consistently with the distribution of the hot and cold gas (see also Fig.~\ref{fig:temp_vlos_180645}), demonstrating the capability of LEM to resolve the multiphase gas in the cluster outskirts clearly and to determine its kinematics (see Section~\ref{sec:results:clumps}). }
\label{fig:phase_space_180645}
\end{figure}

Fig.~\ref{fig:em_hist_180645} illustrates the complexity of the temperature and velocity distributions of the ICM (including all gas phases and substructures) in an individual cluster (cf., Fig.~\ref{fig:temp_pdf2}). The top panel shows the temperature PDF. It follows approximately a Gaussian distribution in the inner region, where the contribution of small substructures to X-ray emissivity is negligible. At larger radii, the distribution becomes more asymmetric and reveals a broad low-temperature tail with $T_{\rm gas}\sim0.1-1\keV$. Outside $0.7r_{200}$, one can see a secondary temperature peak caused by the largest subhalo in the east of the cluster (see Fig.~\ref{fig:map_o7}). We emphasize that this distribution is a mix of the intrinsic ICM temperature distribution (see Section~\ref{sec:sample:temp}) and cold clumps in the cluster outskirts along the LOS. The corresponding shaded regions show the same PDFs after excluding resolved cold clumps using the same mask as in the LEM mock analysis (see Fig.~\ref{fig:map_o7}), highlighting the intrinsic ICM temperature distribution. The thin dotted lines show their best-fit Gaussian distributions, which capture the majority of the PDFs but not the remaining low-temperature tails. These tails contribute a second-order effect to our fitting results (e.g., intrinsic temperature spread of the ICM), which require a more quantitative characterization in future work.
The bottom panel of Fig.~\ref{fig:em_hist_180645} shows the corresponding LOS velocity PDF. The central peak represents the hot ICM component with a peculiar velocity $\simeq-200\kms$, which tends to be slightly broader at a larger radius, reflecting the velocity dispersion of the system. The bumps/peaks on the side show substructures of the cluster. Most of them are infalling towards the cluster centre, which can be clearly seen in Fig.~\ref{fig:phase_space_180645}.

The top panel of Fig.~\ref{fig:phase_space_180645} shows the phase-space distribution of gas cells in the cluster, i.e., radial position vs. radial velocity (see also Fig.~\ref{fig:hist_vr}). The colour encodes the gas temperature. The cyan line shows the (inverse) escape velocity profile $v_{\rm esc}(r)=-\sqrt{2\Phi(r)}$ as a reference, where $\Phi(r)$ is the gravitational potential of the cluster. Inside $r_{500}$ (even $r_{200}$), we barely see any velocity substructures. Most of the cold gas (e.g., $T_{\rm gas}<0.5\keV$) lies outside $\simeq1.2r_{200}$ and has a negative radial velocity close to $v_{\rm esc}$.
In observations, instead of gas radial velocities, LOS velocities are measured. The cluster's phase-space distribution in projection is shown in the bottom panel of Fig.~\ref{fig:phase_space_180645}, i.e., gas projected radial position vs. LOS velocity, using the same set of gas cells as in the top panel. Unlike the top panel, we see a fair amount of cold gas inside $r_{200}$ (even within $r_{500}$). This is a pure projection effect, implying that a non-negligible cold-gas component should always be expected in the X-ray spectra, even extracted from an inner region of a well-relaxed cluster. On the other hand, the same projection effect allows us to probe gas dynamics outside $r_{200}$ in 3D but at projected radii below $r_{200}$, opening a window to gas dynamics in non-virialized regions. The two cyan dashed lines show the maximum escape velocity projected along the LOS, i.e., $v_{\rm esc,max}(r_{\rm proj})=\max{\Big[v_{\rm esc}(r)\sqrt{1-(r_{\rm proj}/r)^2}\Big]}$, where $r\in[r_{\rm proj},\infty)$. They are shifted by $-200\kms$ by taking the peculiar velocity of the cluster into account, enveloping most of the gas points.

\begin{figure}
\centering
\includegraphics[width=0.9\linewidth]{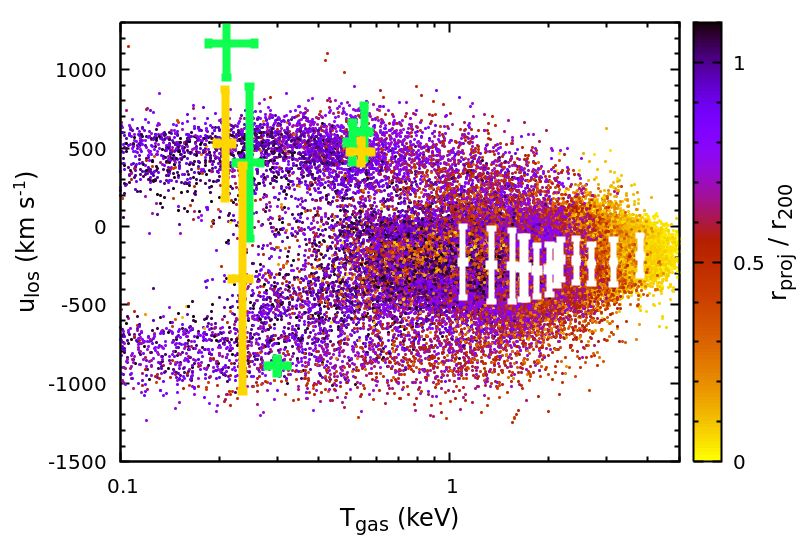}
\caption{Same data as in the bottom panel of Fig.~\ref{fig:phase_space_180645} but showing the gas temperature-LOS velocity diagram. The background points include only the gas cells within $r_{\rm proj}<1.1r_{200}$ (approximately LEM FOV). The colour encodes the projected radius. The cold gas within $(0.7-1)r_{200}$ is recovered as two components with the temperature $0.2-0.3$ and $0.5-0.6\keV$, respectively (see Section~\ref{sec:results:clumps}). }
\label{fig:temp_vlos_180645}
\end{figure}

The overlaid points with error bars in Fig.~\ref{fig:phase_space_180645} show our mock LEM measurements based on the \texttt{1g2b} model. The bars indicate the best-fit velocity dispersion $\kappa\sigma_{\rm los}$, where $\kappa\ (=1.2)$ is a factor included so that the full bar size reflects the full width at half maximum (FWHM) of the velocity field if Gaussian. The white points show the hot ICM component. The cluster's peculiar velocity is accurately recovered. The best-fit FWHM (i.e., the length of the bars) is $\simeq300-400\kms$, consistent with the distributions shown in Fig.~\ref{fig:em_hist_180645}. The green and yellow points indicate the second and third components in our fitting, modelling the cold gas, ordered by their normalization parameter. To exhibit only the most prominent cold clumps resolved in spectra, we show the data points with normalization greater than 5 per cent of the hot component. Their corresponding temperature-LOS velocity diagram is shown in Fig.~\ref{fig:temp_vlos_180645}. Most of these points have positive LOS velocities, in line with the asymmetric velocity PDF shown in Fig.~\ref{fig:em_hist_180645} (solid lines), largely contributed by the prominent substructure in the east of the cluster. Between $0.7-1r_{200}$, the cold clumps appear as two temperature components in our model: the first one with $T_{\rm gas}\simeq0.2-0.3\keV$ and the other with $T_{\rm gas}\simeq0.5-0.6\keV$, reflecting the structures of the temperature PDF in Fig.~\ref{fig:em_hist_180645}. The colder component tends to have larger temperature uncertainties in the fitting. The points showing large velocity dispersions reflect overlapping clumps moving in opposite directions. Note that our simplified model fitting could be improved, especially for resolving multiple cold gas clumps along the LOS, i.e., the points shown in the figure may be further split into more gas components in a more detailed analysis. Additional multi-wavelength observations (e.g., optical, infrared) may also provide prior constraints on the model's parameters, helping to interpret the spectra.

Overall, our mock measurements capture the phase-space distribution of the cold gas and the temperature distribution, demonstrating the capability of LEM to resolve cold clumps spectroscopically and recover their physical properties. The determined $u_{\rm los}$ and $\sigma_{\rm los}$ can provide an approximate radial position of the corresponding cold gas based on the $v_{\rm esc}-v_{\rm esc,max}$ relation. Two important questions to answer for missions like LEM in the future are: what is the minimum radius that the falling cold gas ($\sim0.1-1\keV$) can reach in relaxed clusters? And how does this gas co-exist and co-move (if there is any) with its host galaxies and the hot gas phase?

\subsection{Velocity dispersion and non-thermal pressure} \label{sec:results:vsigma}

Velocity dispersion driven by ICM turbulence can be measured directly through the broadening of X-ray emission lines. In clusters, turbulence dominantly contributes to non-thermal pressure support for atmospheres \citep{Lau2009,Lau2013}, balancing part of the gravitational attraction. For this reason, the hydrostatic mass usually underestimates the true cluster mass \citep[e.g.,][]{Evrard1990,Kay2004,Rasia2006,Nagai2007} and biases the cluster mass function -- the basis of the X-ray cluster cosmology. Direct measurements of the non-thermal pressure are essential to address this issue, correcting cluster mass biases. \textit{Hitomi} mapped the inner region of the Perseus cluster and found an approximate uniform velocity dispersion distribution $\sigma_{\rm los}\simeq150-200\kms$ \citep{Hitomi2016}. \textit{XRISM} will soon observe more clusters, focusing mainly on their brightest central regions. It is, however, non-trivial to extrapolate measurements of $\sigma_{\rm los}$ in the core to the outskirts due to the fact that: (1) AGN feedback and merger-driven sloshing generate strong bulk flows that often dominate the velocity field; and (2) cosmological simulations suggest that the non-thermal pressure fraction is an increasing function of cluster radius, expected to be low near the cluster centre \citep[e.g.,][]{Nelson2014,Shi2015}.
The outskirts of relaxed clusters, alternatively, provide a unique opportunity to measure non-thermal pressure, overcoming both aforementioned issues. This is practically doable only if the instrument simultaneously has both a large grasp and high-spectral resolution.

\begin{figure}
\centering
\includegraphics[width=0.9\linewidth]{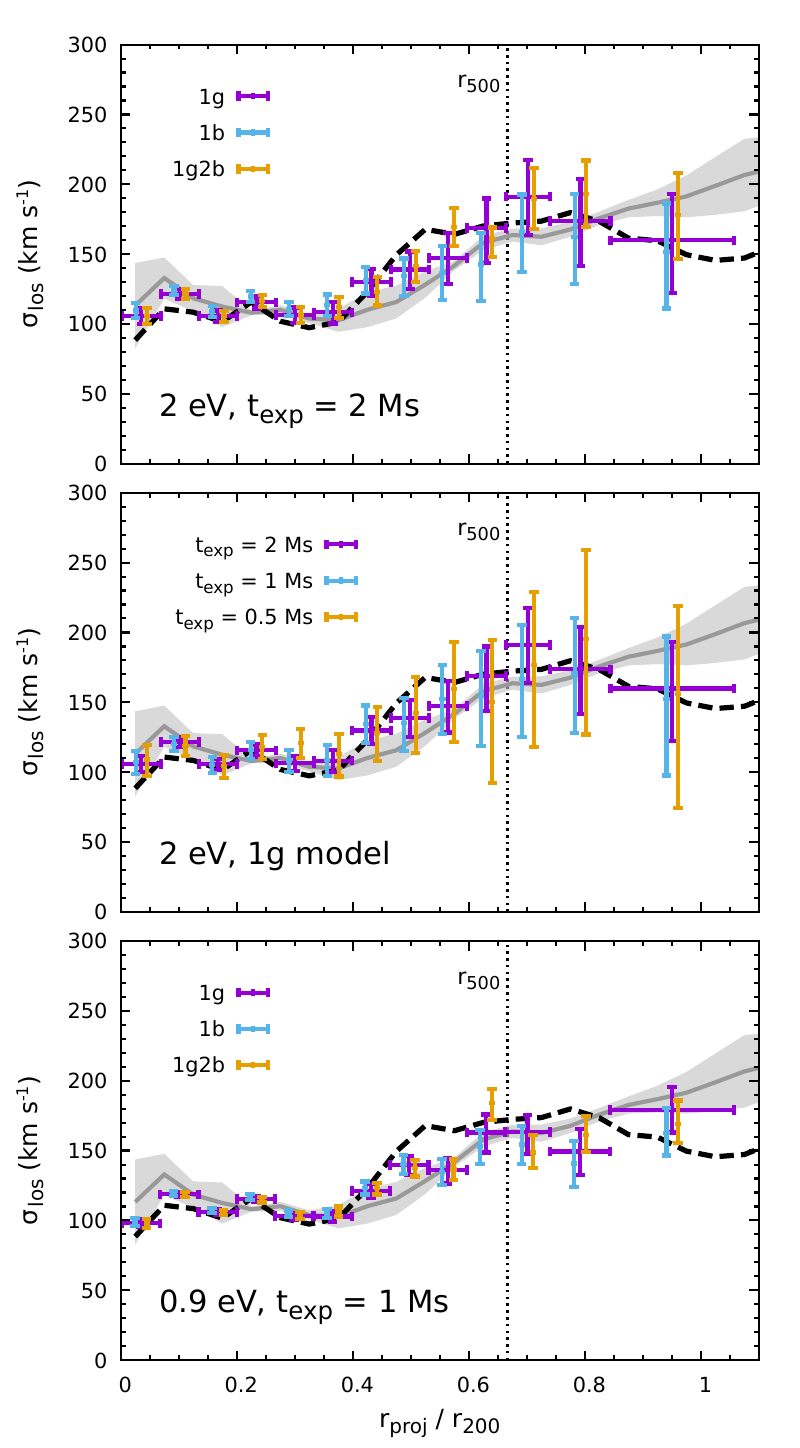}
\caption{Best-fit velocity dispersion profiles of the cluster CL-RM. The grey line and shaded region show the simulated profile of $\sigma_{\rm 1D}$ and its scatter across three independent directions (see Eq.~\ref{eq:sigma1d} and also Fig.~\ref{fig:tng_sigma}). The black dashed line shows the broad-band X-ray-weighted velocity dispersion along the LOS, taking into account only gas cells with $T_{\rm gas}>1\keV$. The three panels compare our mock LEM measurements depending on different fitting models (top), exposures (middle), and energy resolutions (bottom). All the models show consistent results, in line with the simulated expectation. Their errors scale with the exposure $t_{\rm exp}$ as expected. A sub-eV spectral resolution can dramatically improve the velocity dispersion measurements (see Section~\ref{sec:results:vsigma}). }
\label{fig:sigma_180645}
\end{figure}

Fig.~\ref{fig:sigma_180645} shows our experiments on how well LEM can measure $\sigma_{\rm los}$ in the bulk gas component up to $r_{200}$ based on the cluster CL-RM, one of the most relaxed clusters in the TNG simulations. The points show the best-fit velocity dispersion profiles obtained through different fitting models using different exposures and energy resolutions. The grey line and the shaded region show the simulated prediction of $\sigma_{\rm 1D}$ (see Eq.~\ref{eq:sigma1d}) and its scatter in three independent directions ($x,\ y$, and $z$), which implies the small velocity dispersion anisotropy in the cluster. The black dashed line shows the broad-band X-ray-weighted velocity dispersion along the LOS, including only the gas cells hotter than $1\keV$ (see also Fig.~\ref{fig:tng_sigma}). Our fitting schemes/models all provide consistent results, in line with the simulated prediction, shown in the top panel. The \texttt{1g2b} model gives smaller error bars since there are more photons in the data without a mask. Similar results from the \texttt{1g} and \texttt{1b} models imply that the measurement of the velocity dispersion is not very sensitive to the intrinsic temperature structure of the ICM. In the middle panel, we check the dependence of our velocity measurements on the exposure time, $t_{\rm exp}$. We find that the uncertainties in $\sigma_{\rm los}$ scale with an exposure time as $\varpropto1/\sqrt{t_{\rm exp}}$ and $t_{\rm exp}\gtrsim1\Ms$ is necessary to achieve $50\kms$ accuracy in the cluster outskirts ($\ge r_{500}$) depending on the radial binning schemes. In the bottom panel, we experiment with $0.9\eV$ energy resolution while keeping all other instrumental settings the same. The sub-eV resolution turns out to be a game changer. With only half of the exposure ($1\Ms$), the error bars are reduced by a factor of $\simeq2$, equivalent to $\sim8$ times more efficient in determining the velocity dispersion than in $2\eV$ resolution. LEM will have a $0.9\eV$ sub-array in the middle of its detector, which would be useful to advance the measurements, especially in cluster outskirts (e.g., covering a larger region using the same exposure and achieving the same level of velocity precision with a multi-pointing strategy).

\begin{figure}
\centering
\includegraphics[width=0.9\linewidth]{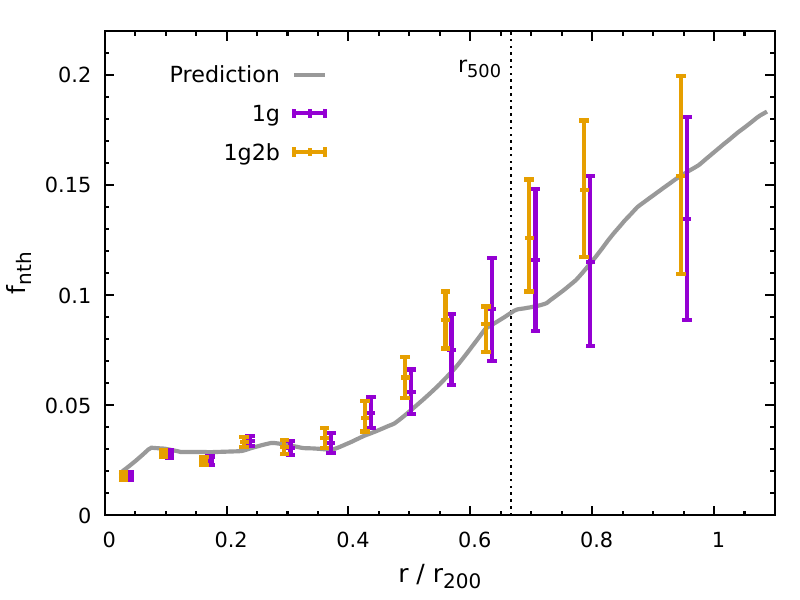}
\caption{Non-thermal pressure fraction profile of the cluster CL-RM. The grey line shows the simulated prediction. The points show the mock LEM measurements, estimated based on the recovered gas density, temperature, and velocity dispersion profiles. They are consistent with the theoretical curve, showing $\sim30$ per cent uncertainties near $r_{200}$. This figure illustrates the LEM's capability of distinguishing low ($\simeq0.1-0.2$) and high ($\gtrsim0.4$) non-thermal pressure fraction within $\simeq r_{500}-r_{200}$ by measuring the most relaxed clusters (see Section~\ref{sec:results:vsigma}). }
\label{fig:fnth_180645}
\end{figure}

Fig.~\ref{fig:fnth_180645} shows the non-thermal pressure fraction profile of the cluster CL-RM, defined as
\be
f_{\rm nth}(r) = \frac{P_{\rm nth}(r)}{P_{\rm gas}(r) + P_{\rm nth}(r)},
\label{eq:fnth}
\ee
where $P_{\rm gas}(r)$ and $P_{\rm nth}(r)\equiv\rho_{\rm gas}(r)\sigma_{\rm 1D}(r)^2$ are the gas thermal and non-thermal pressure profiles, respectively. The grey line shows the simulated prediction, derived from the ICM density, temperature, and velocity dispersion profiles shown in Figs.~\ref{fig:gas_profile}, \ref{fig:tsigma_params}, and \ref{fig:sigma_180645}. The data points show our mock LEM measurements, estimated based on the recovered gas profiles. Their error bars take into account only the velocity dispersion errors, since the statistical uncertainties on recovered density and temperature are negligible by comparison. The $\sigma_{\rm 1D}$ in Eq.~\ref{eq:fnth} is replaced by $\sigma_{\rm los}$, valid for well-relaxed clusters. Both \texttt{1g} and \texttt{1g2b} models recover the prediction, with $\simeq30$ per cent uncertainties near $r_{200}$. The measured $f_{\rm nth}$ can be converted to cluster hydrostatic mass bias ($1-b$) based on cosmological simulations (see, e.g., fig.~B1 in \citealt{Shi2016} for a conversion at $r_{500}$) and/or analytical models \citep[e.g.,][]{Shi2014}. The precision of our measurements is sufficient to distinguish low-level $f_{\rm nth}\simeq0.1-0.2$ within $\simeq r_{500}-r_{200}$, corresponding to the mass bias $1-b\gtrsim0.8$ suggested by the current cosmological simulations \citep[e.g.,][]{Lau2009,Shi2015}, and high-level $f_{\rm nth}\gtrsim0.3$ corresponding to $1-b\lesssim0.7$, considered as a plausible solution for the tension between the cluster X-ray/SZ constraints on cosmological parameters ($\Omega_{\rm m}, \sigma_8$) and those from the primary cosmic microwave background anisotropies \citep{Planck2014,Planck2016}. We note that future weak lensing observations show great potential for calibrating cluster mass scaling relations. However, their mass measurements can still suffer from various systematic effects (see, e.g., \citealt{Umetsu2020} for a review). Direct measurements of cluster hydrostatic masses will be an essential complementary for precision cluster cosmology, as they come with different systematics.

\begin{figure}
\centering
\includegraphics[width=0.9\linewidth]{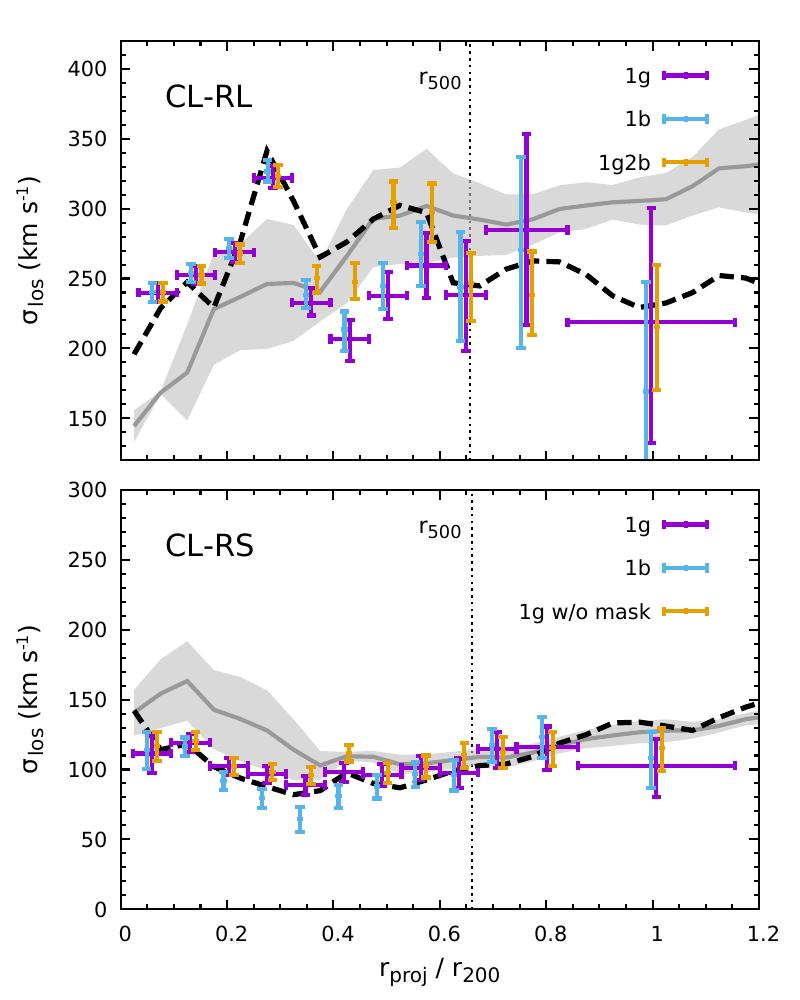}
\caption{Similar to Fig.~\ref{fig:sigma_180645} but for the clusters CL-RL (top) and CL-RS (bottom; see Section~\ref{sec:results:vsigma}). }
\label{fig:sigma_55060}
\end{figure}

\begin{figure*}
\centering
\includegraphics[width=0.9\linewidth]{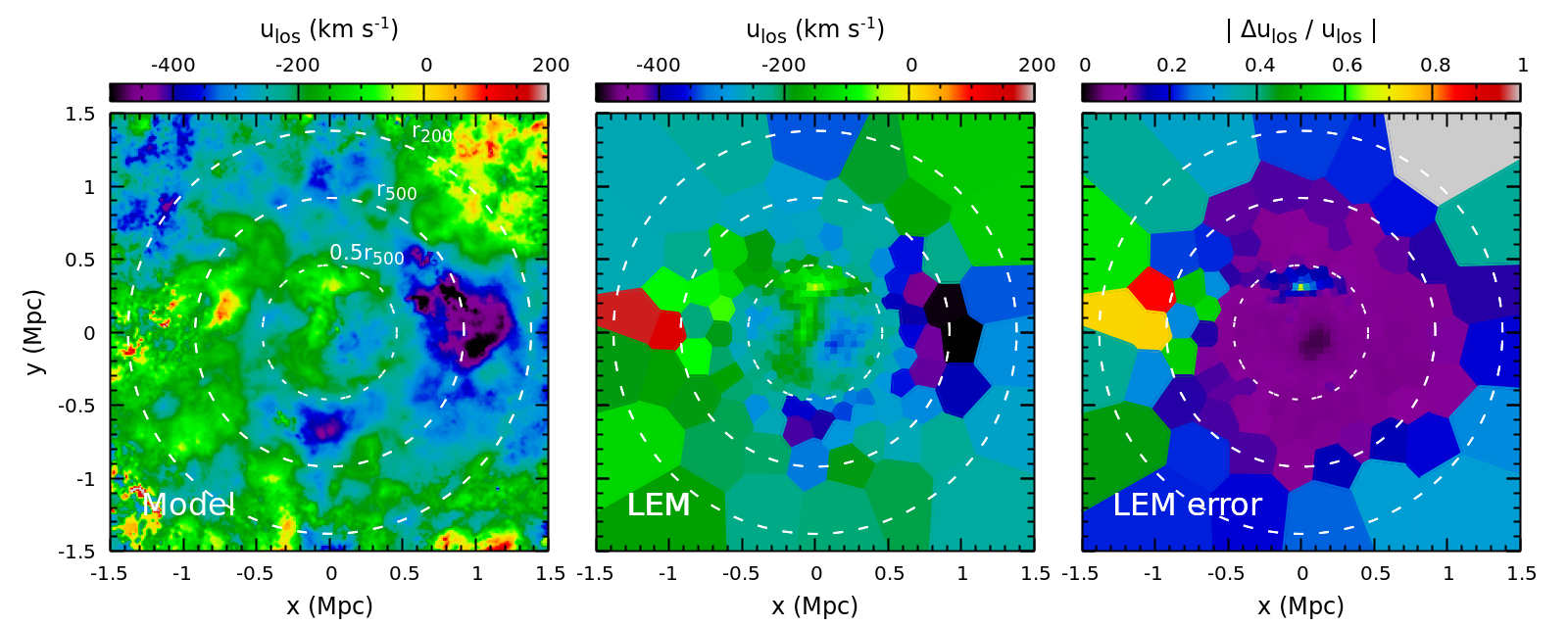}
\caption{Gas LOS velocity distribution of the cluster CL-RM. Three panels from left to right show the simulation (broad-band X-ray weighted), LEM mock measurement, and measurement errors, respectively. In the prediction, we exclude gas cells with $T_{\rm gas}<1\keV$ to remove contributions from the cold clumps. Three white dashed circles indicate characteristic radii. LEM accurately recovers the velocity distribution except for some regions overlapping prominent clumps (e.g., the reddish regions in the middle panel). Within $r_{500}$, we get $\lesssim30$ per cent uncertainties in most of the region (see Section~\ref{sec:results:vlos}). }
\label{fig:map_vlos_180645}
\end{figure*}

In Fig.~\ref{fig:sigma_55060}, we show the velocity dispersion measurements for two other clusters: CL-RL (top), a massive cluster with perturbed gas velocity field at redshift $z=0.12$, and CL-RS (bottom), a quiescent fossil group/cluster at $z=0.06$ (see Table~\ref{tab:cluster_sample}). In the latter, we obtain a good recovery of $\sigma_{\rm 1D}$, similar to Fig.~\ref{fig:sigma_180645}, with even smaller uncertainties everywhere. Because of the absence of prominent substructures within the whole FOV, the spectra from CL-RS can be fitted well with a single gas component. This is confirmed by our \texttt{1g} model, showing similar results with and without a mask to exclude small clumps. The temperature of the cluster lies below $1\keV$ outside $r_{500}$, emitting bright O\,VII and Fe\,XVII lines that provide additional constraints on the velocity dispersion of the hot atmosphere compared to more massive clusters like CL-RM. This explains the small velocity uncertainties in our measurement ($\lesssim20$ per cent near $r_{200}$), suggesting that quiescent fossil groups/clusters are ideal targets for future LEM-type missions to determine the lower limit of $f_{\rm nth}$ in galaxy clusters. The cluster CL-RL, on the contrary, is massive and not fully relaxed. Nevertheless, our measurements are still able to recover $\sigma_{\rm los}$. However, in this cluster, the X-ray-weighted $\sigma_{\rm los}$ profile (dashed curve) noticeably deviates from $\sigma_{\rm 1D}$ (grey region) due to the complexity of the gas substructures (see Fig.~\ref{fig:hist_vr2}). The measured $\sigma_{\rm los}$ does not fully reflect the $f_{\rm nth}$ profile, which implies that future observations should focus on well-relaxed clusters to investigate $f_{\rm nth}$.

\subsection{Line-of-sight bulk motions and velocity structure function} \label{sec:results:vlos}

Many physical questions require statistical information about the velocity field as a function of scale or spatial separation measured through, e.g., power spectrum or structure function. For instance, the slope of the velocity amplitude measured as a function of scale could tell us how the kinetic energy is transported from large to small scales and how it dissipates, providing a window to constrain the microphysical properties of the ICM \citep[e.g.,][]{Gaspari2014,Zhuravleva2019}. By mapping LOS velocities and estimating their velocity structure function, one can constrain the power spectrum of ICM motions, i.e.,
\be
{\rm SF}(d) \propto \int_{0}^{\infty}{P_{\rm 3d}(k)k\dd{k}},
\ee
where ${\rm SF}(d)\equiv\langle|u_{\rm los}(\textbf{r}+\textbf{d})-u_{\rm los}(\textbf{r})|^2\rangle_{\textbf{r}}$ is the second-order structure function, $\langle\rangle_{\textbf{r}}$ indicates an average over all velocity pairs on the sky with a spatial separation $d\ (=|\textbf{d}|)$, and $P_{\rm 3d}(k)$ is the 3D velocity power spectrum (see \citealt{Zuhone2016}, their eqs.~8-12, and also \citealt{Zhuravleva2012}). The injection scales ($\ell_{\rm inj}$) and dissipation ($\ell_{\rm diss}$) scales, key parameters in $P_{\rm 3d}$, define the inertial range. The former is closely related to velocity drivers in the system (i.e., mergers and feedback), and the latter is sensitive to the microphysics of the ICM.

Fig.~\ref{fig:map_vlos_180645} shows our mock LEM results on measuring the LOS velocity distribution of the relaxed cluster CL-RM (middle) and its comparison with the simulated prediction (left). The right panel presents the corresponding measurement error. In the analysis, we adaptively bin X-ray photons with a centroid Voronoi tessellation approach \citep[e.g.,][]{Cappellari2003} by limiting the minimum spatial resolution $l_{\rm min}$ of the bins and the minimum photon number $N_{\rm cnt}$ from the cluster in each bin. In this way, we get roughly uniform grids in the cluster's inner region (e.g., $\lesssim0.5r_{500}$) and a sufficient signal-to-noise ratio in the outskirts. Note that fixing the signal-to-noise ratio over the entire FOV is difficult, given the steep radial profile of the X-ray surface brightness in the cluster outskirts. In this section, we use $l_{\rm min}=0.5'\ (\simeq45\kpc)$ and $N_{\rm cnt}=10^4 {\,\rm cnt\Ms^{-1}}$, and fit the spectra with our standard \texttt{1g2b} model. There are, in principle, a sufficient number of photons to achieve a higher spatial resolution in the cluster's inner regions. To better resolve the hot ICM component, we slightly tighten our prior constraint on its mean temperature parameter ($T_{\rm gas}\in[0.9,\,10\keV]$; see Section~\ref{sec:mock:model}). However, this affects only the regions dominated by cold clumps (reddish regions in the middle panel).

\begin{figure}
\centering
\includegraphics[width=0.9\linewidth]{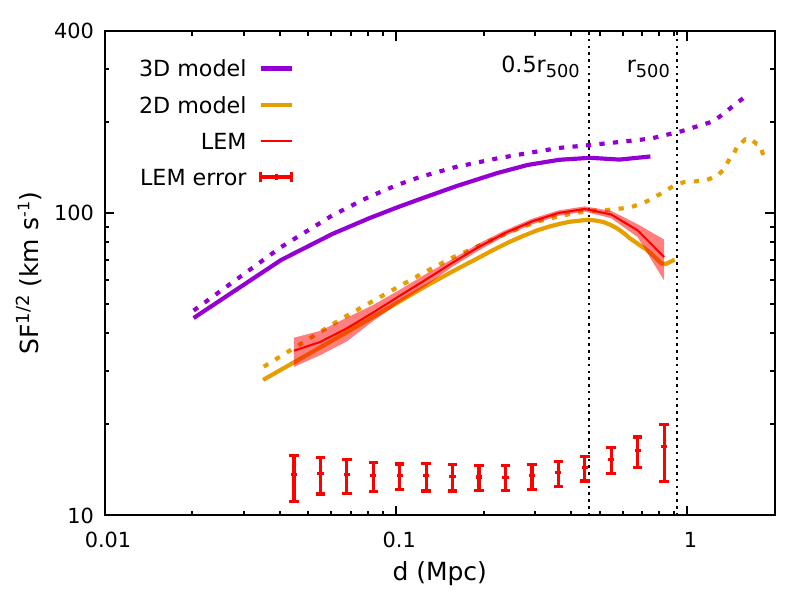}
\caption{Square root of the second-order velocity structure function of the cluster CL-RM. The red line and the shaded region show our LEM mock measurement using all velocity pairs within $0.5r_{500}$ and its uncertainties. The red points with error bars indicate the contribution from the velocity statistical errors (see Eq.~\ref{eq:sf_2comp}). Yellow and purple lines show the corresponding simulation results in 2D and 3D, where the solid and dashed lines are estimated within $0.5r_{500}$ and $r_{500}$, respectively. This figure demonstrates LEM's capability of measuring the ICM velocity structure function and using it to constrain the injection scale (see Section~\ref{sec:results:vlos}). }
\label{fig:sf}
\end{figure}

The middle panel of Fig.~\ref{fig:map_vlos_180645} shows our recovery of $u_{\rm los}$, accurately matching the simulation, particularly within $r_{500}$. Some differences appear in the regions that overlap with the resolved cold clumps (see Fig.~\ref{fig:map_o7}). In our simulation result (the left panel), we exclude gas cells with $T_{\rm gas}<1\keV$ to diminish cold-clump effects. In the LEM mock data, however, it is difficult to resolve hot ICM components from outer regions dominated by clumps (e.g., the big one in the east), even with a tighter prior temperature parameter constraint. We tend to see large velocity scatter and uncertainties in the fitting there. Within $\simeq r_{500}$, no significant clumps are identified. We are able to achieve $\lesssim30\ (20)$ per cent velocity errors (neglecting calibration and gain uncertainties) within $r_{500}\ (0.5r_{500})$, which allows us to constrain the injection scale and a slope of the velocity power spectrum, as discussed below.

Fig.~\ref{fig:sf} shows the square root of the second-order velocity structure function ${\rm SF}(d)$. The red line and the shaded region show the LEM mock measurement and its uncertainties (90 per cent confidence level) using all velocity pairs within $0.5r_{500}$. The uncertainties are modelled using a Monte Carlo approach taking into account (1) the statistical velocity measurement errors, assumed to follow Gaussian distributions (see the right panel in Fig.~\ref{fig:map_vlos_180645}) and (2) the finite spatial size of each bin and its irregular shape. The latter is subdominant in our case, where all bins within $\simeq0.5r_{500}$ have high roundness. In observations, the measured structure function can be written as
\be
{\rm SF}_{\rm obs}(d) = {\rm SF}_{\rm true}(d) + {\rm SF}_{\rm err}(d),
\label{eq:sf_2comp}
\ee
a sum of the true function and the velocity error contribution, if the gas LOS velocity $u_{\rm los}$ and its statistical uncertainty $\Delta u_{\rm los}$ are spatially uncorrelated (see appendix~C in \citealt{Zuhone2016}). The term ${\rm SF}_{\rm err}(d)$ is simply flat if the velocity error is normally distributed with zero mean. In reality, $\Delta u_{\rm los}$ inevitably depends on the radius of the cluster. We model ${\rm SF}_{\rm err}(d)$ in our mock measurement, shown as red points with error bars. It depends only mildly on the separation, explains the slight flattening, and increases uncertainties of the observed structure function at small separations.

The yellow lines in Fig.~\ref{fig:sf} show the projected structure function estimated directly from the simulation data within $0.5r_{500}$ (solid) and $r_{500}$ (dashed). It is not surprising that the LEM result agrees with the simulation, given the well-recovered $u_{\rm los}$ map (Fig.~\ref{fig:map_vlos_180645}). For comparison, we also show the velocity structure functions measured in 3D (purple lines) using the 3D velocity field information rather than the X-ray-weighted projection. The 2D functions reveal smaller amplitudes and steeper slopes due to the strong projection effect. In our cluster sample, the patchiness of the velocity becomes significant outside $\simeq 0.5r_{500}$, largely driven by substructures moving in the ICM. Such bulk motions on $\sim10^2\kpc$ length scales complicate the velocity structure function (compare the solid and dashed lines). For the goal of characterizing turbulence (or random motions) in the ICM, any significant bulk motions should be avoided if possible, e.g., masking regions with prominent LOS bulk motion patterns.

\begin{figure}
\centering
\includegraphics[width=0.9\linewidth]{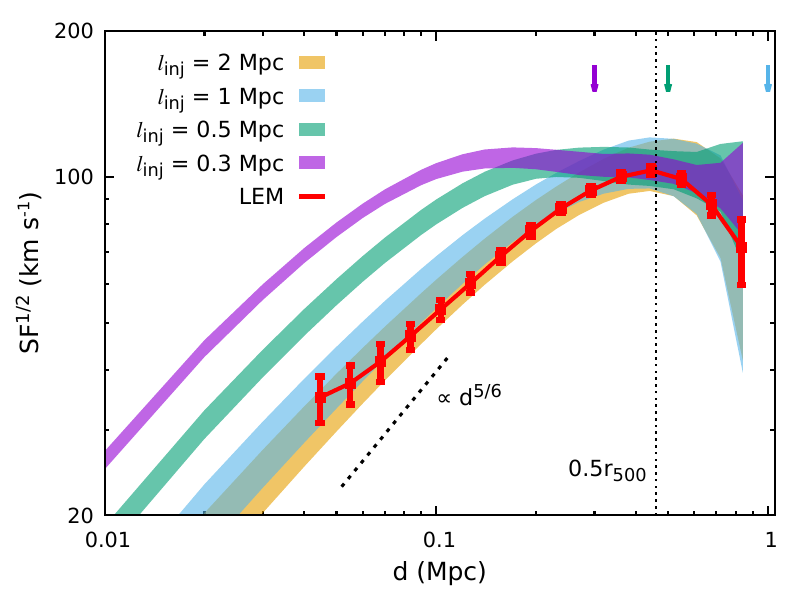}
\caption{Constraining the velocity power spectrum (injection scale and the slope of the inertial range) with LEM's measured structure function. The red line with points shows our LEM mock measurement, the same as the one in Fig.~\ref{fig:sf}. The colour bands show theoretical models with different underlying injection scales ($\ell_{\rm inj}$) indicated by the corresponding vertical arrows, based on Gaussian random velocity fields. The modelled structure functions show a strong dependence on $\ell_{\rm inj}$ if $\ell_{\rm inj}\lesssim1\Mpc$ (approximately the measurement's aperture size $r_{500}$). Our LEM result favours $\ell_{\rm inj}\gtrsim1\Mpc$, demonstrating the robustness of using the velocity structure function to constrain $\ell_{\rm inj}$ of the ICM (see Section~\ref{sec:results:vlos}).  }
\label{fig:sf_gauss}
\end{figure}

To interpret the observed structure function, we generate Gaussian random velocity fields with underlying energy power spectra $E(k)$ as baseline models \citep{Hosking2022},
\be
E(k) = C_0 \frac{(k\ell_{\rm inj})^{\alpha_{\rm e}}}{1+(k\ell_{\rm inj})^{\alpha_{\rm e}-4}}e^{-k\ell_{\rm diss}},
\label{eq:Ek}
\ee
where $C_0$ is a normalization constant setting the standard deviation of $u_{\rm los}$ as $70\kms$ to match our mock LEM structure function's amplitude, $\alpha_{\rm e}=-5/3$ is the spectral index, $\ell_{\rm inj}$ and $\ell_{\rm diss}$ are injection and dissipation scales, respectively. We model the LOS velocity distribution weighted by X-rays as $u_{\rm los}=\int{\varepsilon_{\rm x}u_{z}\dd z}$, where $\varepsilon_{\rm x}=\rho_{\rm gas}^2/\int{\rho_{\rm gas}^2\dd z}$ is the approximate X-ray emissivity. The gas radial density profile of the cluster CL-RM is applied in the model (see Fig.~\ref{fig:gas_profile}). The velocity structure functions are shown as the colour bands in Fig.~\ref{fig:sf_gauss} using the same aperture radius $0.5r_{500}$ as in the LEM mock analysis (red points). The bandwidth shows a standard deviation over 100 realizations, illustrating the level of cosmic variance. We fix $\ell_{\rm diss}=1\kpc$ and find that the modelled structure function exhibits a strong dependence on $\ell_{\rm inj}$ when $\ell_{\rm inj}\lesssim1\Mpc$. Our TNG300 cluster sample favours a large (dominant) injection scale $\ell_{\rm inj}\gtrsim1\Mpc$, and its structure function slope is consistent with Kolmogorov's $5/3$ power law. We stress that flattening of the structure function occurs near $d\simeq\ell_{\rm inj}/3-\ell_{\rm inj}/2$. The aperture size in the measurement (e.g., $r_{500}$ in our case) is thus approximately the upper limit of the injection scale that can be constrained via the velocity structure function. Furthermore, the apparent slope of the function on small scales is shallower than the (asymptotic) theoretical expectation ($\propto d^{5/3}$; see eq.~12 in \citealt{Zuhone2016}), even at $d\simeq\ell_{\rm inj}/10^2$, suggesting that it is important to fit the entire structure function to constrain parameters (e.g., $\ell_{\rm inj}$ and $\alpha_{\rm e}$) while taking into account projection effects, error contributions, and flattening at large separation. We have further tested various $\ell_{\rm diss}$ from $1\kpc$ to $\ell_{\rm inj}/10$ and found that the effect, i.e., slight steepening of the curve, is weak, barely distinguished by LEM and other similar telescopes.

\subsection{Penetrating filaments in galaxy clusters} \label{sec:results:filaments}

The measured gas LOS velocity is contributed by two sources in general: (1) large-scale bulk motions driven by mergers and accretion and (2) volume-filling random motions in the ICM. The latter often dominates the velocity fields in relaxed clusters, as already exemplified in the previous section. Mapping the LOS velocity of unrelaxed clusters, on the other hand, can provide direct evidence of interaction between galaxy clusters and the surrounding cosmic web.

The velocity structures of merging systems can be resolved by observing cluster mergers with their merger axes away from the sky plane. Similar strategies have been applied in the cluster kinematic SZ observations \citep[e.g.,][]{Mroczkowski2012,Adam2017,Sayers2019}. On the other hand, filaments that penetrate into the ICM can also drive Mpc-scale bulk motions \citep{Zinger2016}, playing an important role in thermalizing the IGM. Our TNG300 cluster sample suggests that mergers and deep filament penetration are likely to be often associated with each other. The radial velocities they drive are up to $\sim10^3\kms$, comparable to their free-fall velocities, corresponding to $u_{\rm los}\gtrsim500\kms$ with $\gtrsim30^\circ$ inclination angles.

\begin{figure*}
\centering
\includegraphics[width=0.9\linewidth]{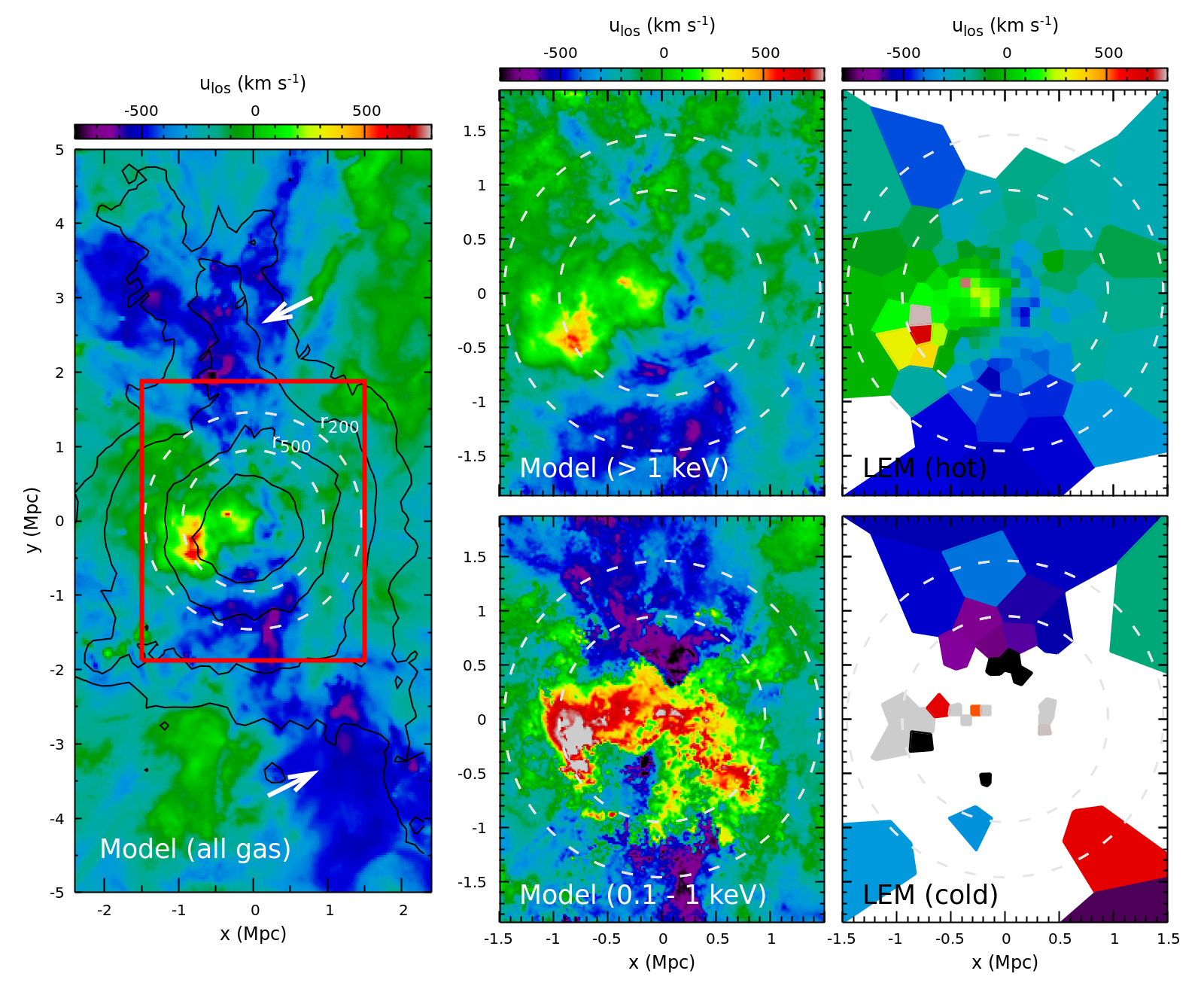}
\caption{Gas LOS velocity distribution of the cluster CL-U -- an unrelaxed cluster with two deeply penetrating filaments indicated by white arrows. The left panel shows the broad-band X-ray-weighted velocity distribution, including all gas cells in the simulation. The overlaid black contours represent the X-ray surface brightness (see also Fig.~\ref{fig:hist_vr}). Zoom-in images of the region marked by the red box are highlighted in the four right panels -- a comparison between the simulations (the middle column) and LEM mock measurements (the right column). The top (bottom) panels compare the LOS velocity of the hot (cold) ICM component, respectively (see more details in the text). Our LEM results well resolve two large-scale filaments associated with the cluster. The north one appears in the hot gas phase, and the south one in the cold phase (see Section~\ref{sec:results:filaments}).  }
\label{fig:map_vlos_150265}
\end{figure*}

The cluster CL-U (see Table~\ref{tab:cluster_sample}) represents such systems, connected with two remarkable filaments from the north and the south, respectively. Fig.~\ref{fig:map_vlos_150265} shows its gas LOS velocity distribution. The left panel exhibits a broad-band X-ray-weighted velocity distribution, including all gas cells in the cluster. The overlaid contours reveal the X-ray surface brightness (see Fig.~\ref{fig:hist_vr}). Both filaments extend beyond the scope of the image and show negative LOS velocities along them (i.e., moving away from the observer). Our mock LEM observations focus on the virial region marked by the red box. To bin X-ray counts, we select $l_{\rm min}=1'$ and $N_{\rm cnt}=2\times10^4 {\rm cnt\Ms^{-1}}$ for our primary goal of identifying large velocity structures in the cluster outskirts. We apply a \texttt{1g1b} model to fit the spectra from the cluster, i.e., one component for the hot gas phase and no more than one for the cold ($N\le1$; see Section~\ref{sec:mock:model}). No additional cold components are included because of the low signal-to-noise ratios at large radii. It is sometimes non-trivial to split cold gas into multiple motion components based on several emission lines (see Section~\ref{sec:results:clumps}). We have compared our results based on the \texttt{1g1b} and \texttt{1g2b} models. They show similar velocity distributions for the hot ICM component. The \texttt{1g2b} model, however, gives larger velocity scatter and larger errors for the cold gas.

The right four panels in Fig.~\ref{fig:map_vlos_150265} show our best-fit LEM results (the right column) and their comparisons with the simulations (the middle column). In the model, we separate the contributions from the different gas phases by using only gas cells with $T_{\rm gas}>1\keV$ (top) and $0.1<T_{\rm gas}<1\keV$ (bottom). Note that the filament in the north appears mostly in the bottom panels, implying its low characteristic temperature, i.e., $\sim5-10$ times lower than the ambient ICM. The filament in the south is revealed in both gas phases in the simulation. The cold phase has a higher LOS velocity than its hot companion; the hot phase shows azimuthally more extended distributions. However, our LEM mock results only recover the latter. In the mock observational results, we compare the best-fit hot and cold components in the fitting model. The top and bottom panels in the right column show the hottest components with the best-fit temperature $T_{\rm gas}>1\keV$ and $T_{\rm gas}<1\keV$, respectively. The latter represents the cold gas phase, since the hot ICM cannot always be resolved from the cluster outskirts (i.e., two components both capture the cold gas in the fitting). We present only the bins with velocity errors $|\Delta u_{\rm los}|<200\kms$, i.e., non-empty regions in the figure.
The LEM images show similar velocity patterns as in the simulation. Two filaments are well resolved not only in their LOS velocities but also their morphologies in projection (e.g., gaseous boundaries, opening angles), which are important to explore the co-evolution of multiphase gas and member galaxies inside the filaments.

Filaments connected with a cluster can lead to anisotropic mass accretion and reshape ICM boundaries \citep{Zhang2021,Vurm2023}. Their interactions with the ICM are still poorly resolved in the current cosmological simulations, e.g., the formation of boundary layers and shock structures. This is why we focus only on detecting large-scale features of the penetrating filaments in this section (e.g., gas velocity and geometry). In reality, the complexity of soft X-ray lines can be vital diagnostics of the instabilities and mixing formed near the filament-ICM interfaces. Filaments inside clusters are typically cold (i.e., $0.1-1\keV$), unable to produce Fe-K lines. LEM-type X-ray telescopes provide unique capabilities to explore their co-evolution with the clusters.

\section{Conclusions} \label{sec:conclusions}

High-resolution X-ray spectroscopy in soft X-rays will open a new window to map multiphase gas in galaxy clusters, especially in the cluster outskirts, allowing us to probe the transition between halos' virialized and non-virialized regions. Using LEM as an example, we illustrate the essential role of eV-level spectral resolution in separating cluster emission from the Galactic foreground. It enables us to probe the strongest lines (e.g., O\,VII, O\,VIII, and Fe-L complex) from the ICM and cold substructures and to measure gas velocities and metal abundances from line shifts, line broadening, and line ratios. Prominent line features and their dependence on gas temperature guarantee that they are sensitive tracers of gas in different phases. Cluster targets must be within specific redshift ranges, ideally $z\simeq0.06-0.1$.

By mocking LEM observations with full background information, analysing their images/spectra, and comparing extensively with simulation predictions, we demonstrate that, with $\simeq1-2\Ms$ exposures, LEM can reliably probe the gas thermodynamic, chemical and kinetic properties of the ICM from the cluster core to the outskirts. We emphasize that our methodology and main findings are applicable to other similar X-ray missions as well, e.g., HUBS and Super\,DIOS, providing theoretical guides for their development of the ICM/IGM science objectives.
Our main conclusions are summarized below.
\begin{itemize}
  \item Gas density and temperature radial profiles of the bulk, volume-filling ICM can be measured with high accuracy out to $r_{200}$ and beyond, as well as the width of the Gaussian-shape intrinsic temperature distribution. They are not prominently affected by gas clumpiness. Both the mean temperature and the normalized variation ($\sigma_{\rm temp}\simeq0.1-0.3$) are essential parameters to describe the ICM temperature. The latter provides a unique opportunity to constrain physical processes in galaxy clusters (e.g., turbulence, mixing, and feedback). Our measurements are robust to the methods used to account for cold clumps when fitting spectra (see Section~\ref{sec:results:gas}).
  \item Iron and oxygen abundances of the ICM can be accurately recovered in our experiments up to $r_{200}$. To avoid any biases in measuring these abundances (Fe in particular), it is important to assume a temperature distribution of the ICM with a specific non-zero width in the fitting model (see Section~\ref{sec:results:metal}).
  \item Multiple gas phases, particularly prominent in the cluster outskirts, can be separated spectroscopically with a spectral resolution similar to that of LEM by measuring the velocities of different gas components along the LOS. It allows us to map the gas phase-space distribution, providing valuable information on the assembly history of galaxy clusters, star formation quenching, and transport properties of the ICM. In particular, thanks to projection effects, cold gas measured within $r_{200}$ (even $r_{500}$) provides a window to probe gas dynamics in non-virialized regions (see Section~\ref{sec:results:clumps}).
  \item We measure gas velocity dispersion profiles with $\leq50\kms$ uncertainties in well-relaxed clusters and estimate the corresponding non-thermal pressure fraction. We demonstrate LEM's capability in distinguishing low non-thermal pressure fraction $f_{\rm nth}\simeq0.1-0.2$ suggested by the current cosmological simulations and high $f_{\rm nth}\gtrsim0.4$, considered as a plausible solution for the ($\Omega_{\rm m},\ \sigma_8$) tension in cluster cosmology (see Section~\ref{sec:results:vsigma}).
  \item We map the gas LOS velocity distribution of a well-relaxed cluster with $\lesssim30\,(20)$ per cent uncertainties within $r_{500}\,(0.5r_{500})$ and use it to estimate the 2nd-order velocity structure function. The projected structure function is steeper than the 3D one, whose shape is sensitive to the dominant injection scale of the system if it is smaller than the aperture size of the observation. Our TNG300 cluster shows consistency with Kolmogorov's $5/3$ turbulence power law in 3D and favours a large injection scale $\ell_{\rm diss}\gtrsim1\Mpc$, implying LEM's feasibility to constrain properties of ICM turbulence (see Section~\ref{sec:results:vlos}).
  \item We identify two Mpc-scale penetrating filaments inside a TNG300 unrelaxed cluster in our mock LEM observation. We resolve their spatial morphology, e.g., gaseous boundaries in projection and opening angles, and LOS velocities (both cold and hot components). Such observation will open a new windows in the future to investigate ICM-filament interactions, e.g., boundary layers, mixing, and stripping of the gas (see Section~\ref{sec:results:filaments}).
\end{itemize}

In this study, we provide an example of bridging numerical simulations and X-ray observations self-consistently. Instrumental response, observational background, and modelling mock data with observational approaches are among the important steps to be taken for correct predictions from numerical simulations testable with observations. On the one hand, cosmological simulations provide insights into developing observational projects and interpreting observational data. For example,
\begin{itemize}
  \item We show that the dynamical state of galaxy clusters has strong effects on various X-ray observables. Only well-relaxed clusters can be used to constrain the non-thermal pressure fraction of the ICM (see Section~\ref{sec:sample:vsigma}). In simulations, we show that estimating the gas radial velocity histogram in radial shells provides a robust way to classify clusters into different dynamical states. The TNG300 simulation also suggests that deep filament penetration into the ICM is often associated with the merger process, which will guide the selection of cluster targets in future observations (see Section~\ref{sec:sample:vr}).
  \item We characterize the PDFs of temperature fluctuations of the ICM in TNG300 clusters. They can be well described by log-normal distributions in radial shells and Gaussian distributions with some skewness in projected 2D annuli (see Section~\ref{sec:sample:temp}). These results motivate our \texttt{gaussbvapec} model applied in the mock observational analysis. Furthermore, we show that the resolved cold clumps ($\simeq0.1-1\keV$) are widely spread not only in the cluster outskirts but also within $r_{500}$ even in well-relaxed clusters due to the projection effect. These clumps must be resolved spatially (see Section~\ref{sec:mock:mask}) and/or spectroscopically (see Section~\ref{sec:mock:model}) to take their effects into account.
\end{itemize}
On the other hand, we emphasize that some physical processes might not be properly resolved or might even be absent (e.g., small-scale turbulence, mixing layers) in the current cosmological simulations, which limits our discussion on their properties in this study. Future observations will open a large discovery space in these areas and stimulate the development of more physically-motivated theoretical models. Moreover, the present study assumes a perfect understanding of the atomic database and sky background. However, these are non-trivial in reality, especially for high-resolution X-ray spectroscopy. In the coming decade, we will benefit from the missions including \textit{XRISM} and SRG/eROSITA in preparation for the era of LEM, HUBS, and Super\,DIOS.

\appendix

\section{Supplementary figures} \label{sec:appendix}

\begin{figure}
\centering
\includegraphics[width=0.9\linewidth]{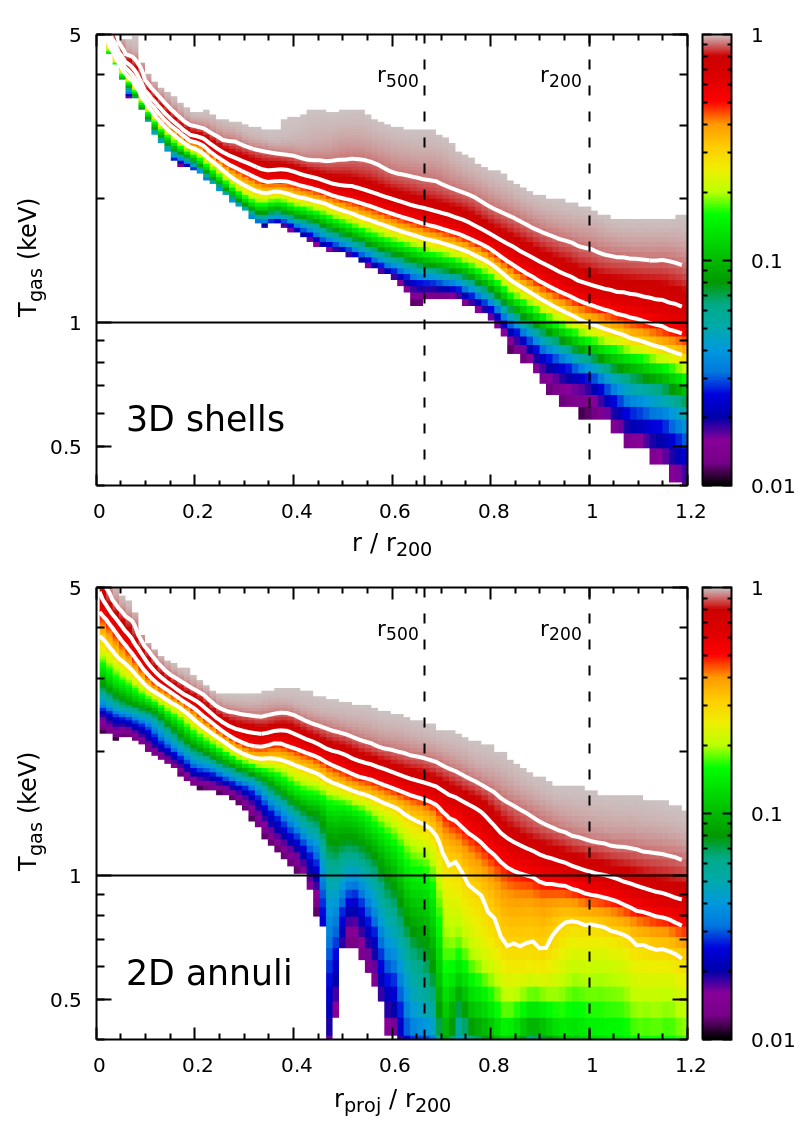}
\caption{Normalized cumulative distribution of the gas temperature in radial shells weighted by gas mass (top) and in 2D annuli along the LOS weighted by broad-band X-ray emissivity (bottom) for the cluster CL-RM. The colour shows only the range of $0.01-0.99$. The white contours indicate 0.3, 0.5, 0.7, and 0.9. This figure illustrates the robustness of using a simple $T_{\rm gas}>1\keV$ criterion to exclude cold clumps along the LOS within $r_{200}$ (see Appendix~\ref{sec:appendix}). }
\label{fig:thist_cumu_180645}
\end{figure}

\begin{figure*}
\centering
\includegraphics[width=0.9\linewidth]{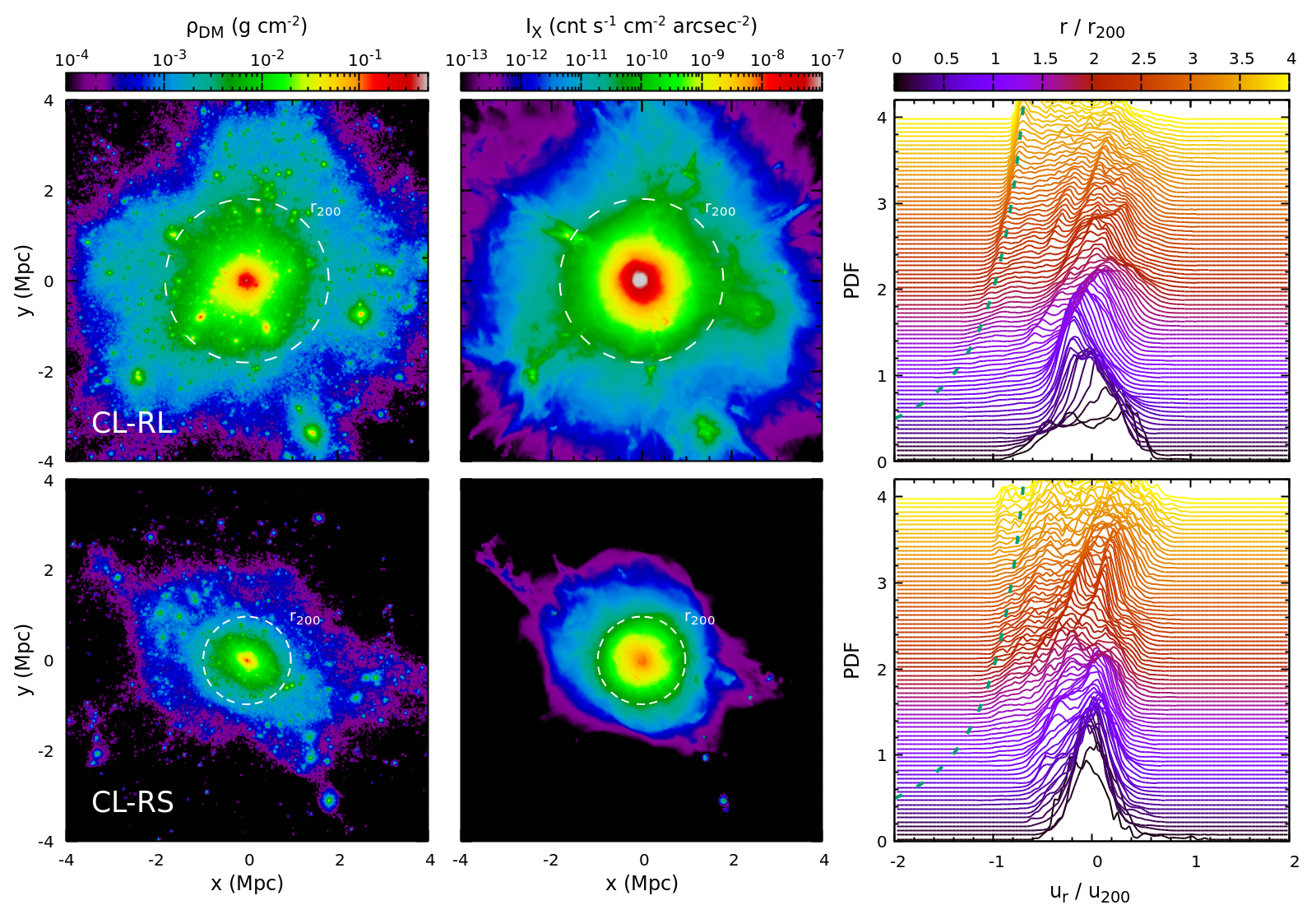}
\caption{Similar to Fig.~\ref{fig:hist_vr} but for the clusters CL-RL (top) and CL-RS (bottom). CL-RL is a massive one and not yet fully relaxed. Despite lacking prominent substructures, its radial velocity histogram is noticeably asymmetric within $r_{200}$. CL-RS is a quiescent fossil group/cluster, isolated from any other large-scale structures (e.g., filaments and halos). It provides an ideal opportunity to constrain the lower limit of non-thermal pressure fraction in clusters (see Appendix~\ref{sec:appendix}).  }
\label{fig:hist_vr2}
\end{figure*}

Fig.~\ref{fig:thist_cumu_180645} shows the cluster CL-RM's normalized cumulative distribution of the gas temperature in radial shells weighted by gas mass (top) and in 2D annuli weighted by broad-band X-ray emissivity (bottom). Near and outside $r_{500}$, a large amount of X-ray flux is contributed by cold clumps and diffuse low-temperature gas at large cluster radii along the LOS. In this study, we use a uniform temperature criterion (i.e., $T_{\rm gas}>1\keV$) to exclude contributions of cold clumps in the simulated predictions (see, e.g., Figs.~\ref{fig:tng_sigma}, \ref{fig:gas_profile}, and \ref{fig:tsigma_params}). Fig.~\ref{fig:thist_cumu_180645} shows that $\lesssim1\ (30)$ per cent gas mass is colder than $1\keV$ at $r_{500}\ (r_{200})$. However, along the LOS, the gas contributes $\simeq20\ (70)$ per cent of the X-ray flux with $T_{\rm gas}<1\keV$. Given the steep X-ray surface brightness radial profile of the ICM, our approach is simple and robust. However, we note that, due to the broad ICM-temperature distribution in the cluster outskirts, a small fraction of the ICM with moderate temperature is inevitably excluded in our estimation. This does not significantly affect our results except for the predicted X-ray-weighted temperature (see Fig.~\ref{fig:tsigma_params}).

Fig.~\ref{fig:hist_vr2} shows the dark matter, X-ray surface brightness, and gas radial velocity distributions of the clusters CL-RL and CL-RS, similar to Fig.~\ref{fig:hist_vr}.

\section*{Acknowledgments}

The material is based upon work supported by NASA under award number 80GSFC21M0002.
Part of the simulations presented in this paper were carried out using the Midway computing cluster provided by the University of Chicago Research Computing Center. IZ is partially supported by a Clare Boothe Luce Professorship from the Henry Luce Foundation.
SE acknowledges the financial contribution from the contracts ASI-INAF Athena 2019-27-HH.0, ``Attivit\`a di Studio per la comunit\`a scientifica di Astrofisica delle Alte Energie e Fisica Astroparticellare’' (Accordo Attuativo ASI-INAF n. 2017-14-H.0), and from the European Union’s Horizon 2020 Programme under the AHEAD2020 project (grant agreement n. 871158).
D. Nelson acknowledges funding from the Deutsche Forschungsgemeinschaft (DFG) through an Emmy Noether Research Group (grant number NE 2441/1-1).
D. Nagai is supported by NSF (AST-2206055 \& 2307280) and NASA (80NSSC22K0821 \& TM3-24007X) grants.
WF acknowledges support from the Smithsonian Institution, the Chandra High Resolution
Camera Project through NASA contract NAS8-03060, and NASA Grants 80NSSC19K0116, GO1-22132X, and
GO9-20109X.

\section*{Data Availability}

The data underlying this article will be shared on reasonable request to the corresponding author. The IllustrisTNG simulations, including TNG300, are publicly available at \url{www.tng-project.org/data} \citep{Nelson2019}. The LEM instrumental responses are available at \url{https://www.lem-observatory.org} \citep{Kraft2022}.

\bsp	
\label{lastpage}
\end{document}